\documentclass[12pt,a4paper]{report}
\usepackage[cp1250]{inputenc}
\usepackage{wrapfig}

\usepackage{amsmath}
\usepackage{graphicx}
\usepackage{amssymb}
\usepackage{epsfig}

\newcommand{\titl}[1]{{\centering\Large\bf #1\par}\bigskip}
\newcommand{\name}[1]{{\centering\rm\normalsize #1\par}\bigskip}

\newcommand{\adr}[1]{{\it \normalsize #1\par}\medskip}

\begin{document}
\begin{titlepage}
\begin{center}

\includegraphics[width=0.8\textwidth]{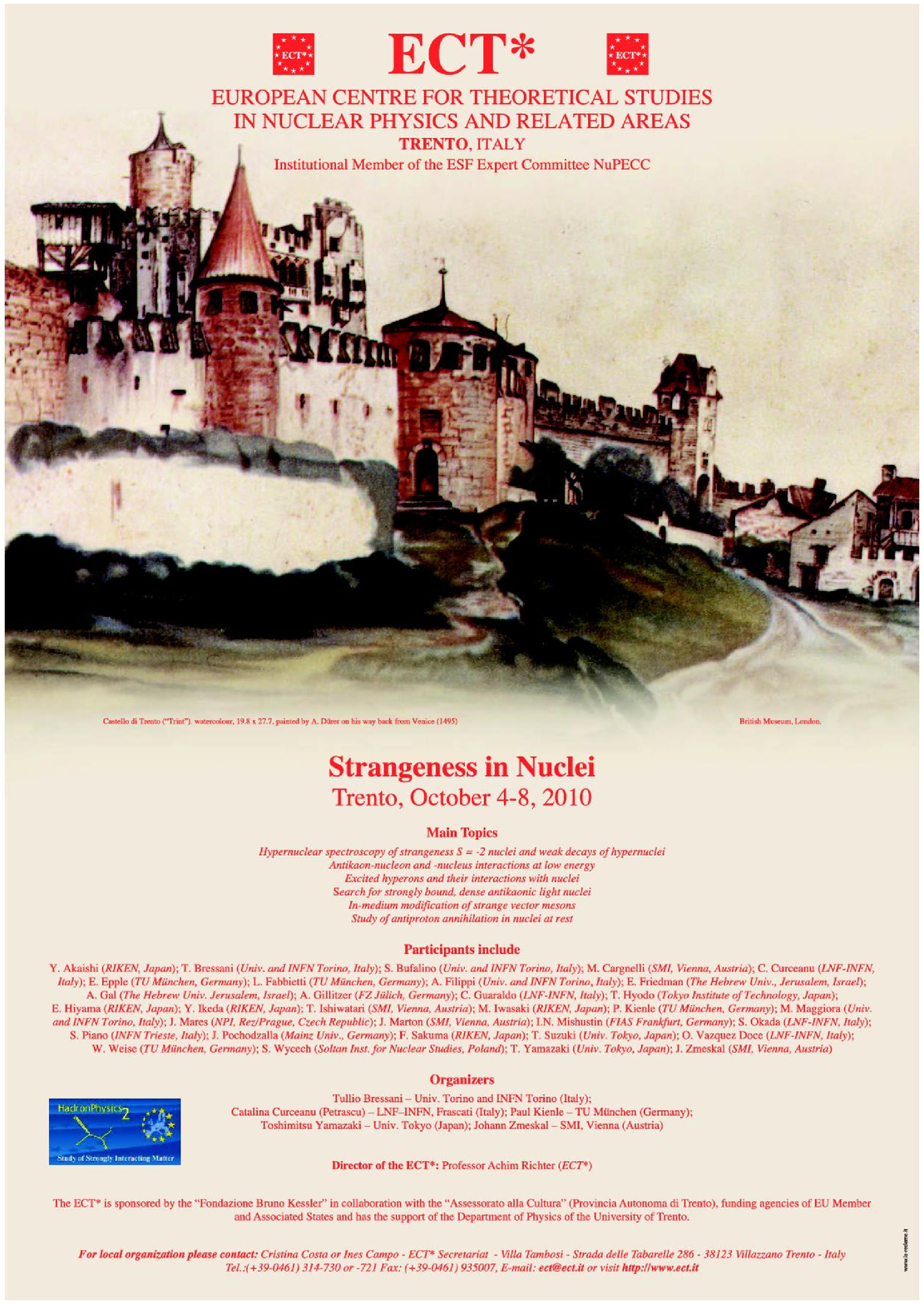}\\[1.5cm]
\textsc{\Large Mini-Proceedings}\\[0.5cm]
\textsc{\Large ECT* Workshop}\\[1.5cm]
\textsc{\LARGE Strangeness in Nuclei}\\[3cm]

\text{}
\text{Eds. C. Curceanu (INFN-LNF/Frascati) 
and J. Zmeskal (SMI/Vienna)}\\[1.5cm]

\end{center}
\end{titlepage}

\tableofcontents
 
\clearpage

\addcontentsline{toc}{section}{
{\bf Introduction} \\
C.~Curceanu, J. Zmeskal}

%





\titl{Introduction}

\name{C.~Curceanu$^{1}$, J. Zmeskal$^{2}$
}

\begin{center}
\adr{
$^1$ INFN, Laboratori Nazionali di Frascati, Frascati (Roma), Italy \\
$^2$ Stefan-Meyer-Institut f\"{u}r subatomare Physik, Vienna, Austria\\
}
\end{center}
\noindent
{\bf MAIN TOPICS}\\
This workshop brought together international experts in the research area of strangeness 
in nuclei physics, working on theory as well as on experiments, to discuss the present status, 
to develop new methods of analysis and to have the opportunity for brainstorming towards 
future studies, going towards a deeper understanding of the hot topics in the low-energy QCD 
in the strangeness sector.

Main topics of discussion were:
\begin{itemize}
\item 	Hadronic atoms
\item	Hypernuclear physics
\item	Meson (in particular kaon)-nucleon scattering status
\item	Low-energy effective theories 
\item	In-medium modification of vector mesons
\item	Excited hyperons and their interaction with nuclei
\item	Deeply bound meson-nuclear states: theoretical status
\item	Antiprotons as hadronic probes
\item	EU HadronicPhysics FP7 LEANNIS Network and its future
\item	Experimental results:
	\begin{itemize}
	\item DEAR and SIDDHARTA at DAFNE
	\item Kaonic Helium by E570 at KEK
	\item Deeply bound mesonic nuclear states:
	\begin{itemize}
	\item   E549  at KEK
	\item   FOPI and HADES at GSI
	\item   FINUDA and KLOE at DAFNE
	\item   OBELIX results
	\item   Dubna  results
	\item   DISTO at Saturne
	\end{itemize}
	\end{itemize}
\item	Next-generation experiments
	\begin{itemize}
	\item	Experiments at GSI: future of FOPI and HADES
	\item	Experiments at  DAFNE: SIDDHARTA2, AMADEUS and future plans; 
	\item	E15 and E17 at J-PARC
	\item	Facility for antiproton annihilation in nuclei physics: JPARC and CERN
	\end{itemize}
\end{itemize}

\noindent
{\bf SCIENTIFIC REPORT}

The strangeness nuclear physics is an extremely valuable tool for studying fundamental 
interactions and symmetries in a fairly direct way, complementing the high energy physics 
studies performed at LHC and elsewhere. Important information regarding the low-energy, 
non-perturbative, regime of QCD can be gained from these types of research. Since the 
pioneering days, decades ago, new technologies (in accelerators and detectors) were developed
 in order to perform precision experimental studies to clarify open issues (such as the still
 existing discrepancies between experiment and theory in  kaonic atoms, the quest for kaonic 
nuclear clusters or studies on hypernuclei with $S=-2$) and to extract new data with unprecedented
 accuracy. One can say that the field is experiencing a happy coincidence in which the progress 
achieved in accelerator physics is paralleled by the advances in detector physics.  
Theory has, meanwhile, performed equally important steps forward towards a deeper 
understanding of the involved physics processes.

The Workshop was a continuation, a deepening and an enlargement of physics sectors of 
the very successful 2006 ECT* Workshop on ``Exotic hadronic atoms, deeply bound kaonic 
nuclear states and antihydrogen'' and of the 2009 ECT* Workshop  ``Hadronic atoms and 
nuclei - solved puzzles, open problems and future challenges in theory and experiment''.

The present Workshop brought together international experts in the research area 
of strangeness in nuclei physics, working on theory as well as on experiments, to 
discuss the present status and the recent important progress, to develop new methods
of analysis and to have the opportunity for brainstorming towards future studies.

Going more into detail, the following main items were discussed:
\begin{enumerate}
\item  {\bf Hypernuclear spectroscopy, including the strangeness S = -2 nuclei and weak decays
 of hypernuclei} which were new topics and refer to experimental results already existent
 (as FINUDA) and to experiments proposed for the J-PARC and FAIR facilities for the 
search for double $\Lambda$ nuclei and the synthesis of cascade nuclei and their weak 
decays including their theoretical relevance for the strangeness interaction in nuclei. 

\item {\bf Antikaon-nucleon and -nucleus interactions at low energy} for which we had new 
results from the analysis of the kaonic hydrogen, deuterium and $^3$He data from SIDDHARTA at LNF. 
This new data were at the basis of a very fundamental discussion of the complex nature of the 
antikaon-nucleon interaction governed by hyperon resonances close to threshold. Future of 
this sector will see the SIDDHARTA2 and E17 experiments in data taking in the coming years, 
the first one at DA$\Phi$NE the second at J-PARC. These experiments plan to go to perform even
 more precise kaonic atoms measurements for more atoms, as required by theoreticians working in the field.

\item {\bf Excited hyperons and their interactions with nuclei}, especially the double 
pole property of the $\Lambda(1405/1420)$ resonance which is crucial for understanding of 
our central theme ``Strangeness in Nuclei''. New data for this much discussed topic 
were shown from HADES at GSI and they gave a key database for a theoretical understanding 
of the strong attraction mediated by antikaons in nuclei.

\item {\bf Search for strongly bound, dense antikaonic light nuclei} 
predicted by Akaishi and Yamazaki in 2002 were strongly debated at the present workshop where 
final result of DISTO data analysis were presented and discussed. In addition, FOPI preliminary 
results from the recent experiment on the search for the $K^-pp$ state in the $pp$ reaction at 3.1 GeV 
bombarding energy were introduced. Furthermore final results were reported on $\Lambda-n$ and 
$\Lambda-p$ correlations following stopped $K^-$ absorption at rest in $^4$He measured in the E549 
experiment which give indications of deeply bound tri-baryon systems. E15 at J-PARC is going on 
line for a search of the K-pp state using the in flight $K^-(^3{\rm He}, n)$ reaction with a 
kinematically complete experimental set up - the status of this experiment was discussed. 
AMADEUS future dedicated experiments for these type of studies at DA$\Phi$NE were as well discussed.

\item {\bf In-medium modification of strange vector mesons}, such as $\phi$-mesons with hidden 
strangeness, has been studied recently using invariant mass spectroscopy in the $p+A$ reaction by 
the KEK-PS experiment E325 to determine its mass shift and width. For the study of 
bound states of $\phi$-mesons in nuclei, missing mass spectroscopy using antiproton annihilation 
reactions with small momentum transfer have been proposed recently by Iwasaki {\it et al.} to 
be performed with antiproton beams at J-PARC and at a later stage at FAIR. The status of 
the proposal was reviewed and new ideas discussed and implemented.

\item {\bf Study of antiproton annihilation in nuclei at rest} using a $4\pi$-detector system with 
the capability of detecting and identifying all charged and neutral particles and measuring their 
four momentum in order to study the reaction mechanisms with high statistics and kinematically 
complete was proposed. A search for double antikaonic strongly bound and dense nuclear systems, 
and systems of high energy density produced by antiproton annihilation in nuclei will be proposed at J-PARC.

\end{enumerate}

Moreover, in the framework of the present Workshop two other activities were organized:
\begin{enumerate}
\item	presentation and discussion of the EU FP7 LEANNIS Network in the framework of 
HadronPhysics2 program and its continuation in HP3

\item	physics conference ``La Fisica moderna tra arte e medicina'' by Catalina Curceanu 
at the Galile Galilei High School of Trento
 (http:\slash\slash{}trentinocorrierealpi.gelocal.it\slash{}cronaca\slash{}2010\slash{}10\slash{}11\slash{}news
\slash{}la-fisica-moderna-tra-arte-e-medicina-2509188)
\end{enumerate}

\noindent
{\bf RESULT AND HIGHLIGHTS}

The field of Strangeness in Nuclei is a very active one, as was fully proven 
during this Workshop. On one side there are many new and important experimental results 
coming from precision experiments performed or undergoing in GSI, DAFNE, KEK, or 
from re-analyses of existent older data from OBELIX, DISTO or JINR experiments, just to name a few. 
On the other side, theoretical tools have performed important steps forward, motivated by the 
new coming results: not only at technical level, but, even more important, on the understanding 
of underlying physical processes or questions to be explored in the upcoming experiments. 

There are many solved problems, as kaonic hydrogen and kaonic helium measurements - which 
were understood, both due to the newer experiments (as SIDDHARTA at DA$\Phi$NE) and to theoretical 
interpretation, but many open problems are still present. Actually, the number of open problems 
is increasing, theories are dealing with one or two  $\Lambda(1405)$, with various potential models below the 
threshold on kaon-nucleon interaction, having as consequence the possibility or not to create deeply 
bound kaonic nuclear states. The sector of hypernuclear physics is still long way to go, especially 
in going towards S=-2 sector. Important questions were targeted and formulated, but they still 
need both experimental results and deeper theoretical understanding. The future challenges 
in the sector of strangeness in nuclei physics are many and were, for the first time, focussed and 
formulated in a unitary framework. 

One can rightly say that we can look forward to the future of low-energy QCD with strange quarks 
as a quantitative, precision science.
This Workshop definitely gave an important contribution in this sector and we plan to apply for a 
future one eventually in 2012 when more experimental data and theoretical work will become available.

During the workshop few actions, apart of regular talks given both by theoretician experts in 
the field and experimentalists, were organized. The EU FP7 framework programs was presented:
in particular a discussion was dedicated to the LEANNIS Network (HadronPhysics2 FP7) by J. Marton, 
just dealing with physics discussed in the present workshop; future important contacts and modalities 
to optimize future contacts were discussed and some decision taken (for example related to the 
continuation and renewal of this Network in HP3).  Moreover, a physics conference "La Fisica moderna 
tra arte e medicina" by Catalina Curceanu at the Galileo Galilei High School of Trento was organized 
with the support of ECT* (in particular of Cristina Costa).

The workshop gathered together world-leading experimental and theoretical experts in the field and 
young scientists and students, providing a state-of-the-art overview of the field of hadronic atoms
 and kaonic nuclear states. The young participant's percentage was about 50\%, which is one of 
the successes of the Workshop.
Moreover, participants from many countries took part, making it an occasion not only of scientific 
exchange, but of cultural and social ones too, proving once again scientists are part of society, 
and their role is an important one.

The future of the field looks bright and promising - in good health, with an ideal mixing of 
experts and young, theoreticians and experimentalists, understood items and puzzles. 
The organization of these type of Workshops in the ideal environment of ECT* contributes to 
the progress of the field. Since many experiments are just going or are about to start, 
and since theories and methods are evolving in parallel, we plan to organize other workshops 
in this field in the coming years, at ECT*.

Last but not least a note of merit: the organization of the Workshop by ECT* 
(special thanks to the ECT* Director, Prof. A. Richter, and to Ines Campo, the Workshop secretary) 
was excellent.



 
\setcounter{equation}{0} 
\setcounter{figure}{0}
\clearpage

\addcontentsline{toc}{section}{
{\bf Formation of few-body kaonic nuclear clusters} \\
 Y. Akaishi, J. Esmaili, T. Yamazaki}

%





\titl{Formation of few-body kaonic nuclear clusters}

\name{ Y. Akaishi$^{1,2}$, 
J. Esmaili$^{1,3}$, T. Yamazaki$^{1,4}$
}

\adr{
$^1$ RIKEN Nishina Center, Wako, Saitama 351-0198, Japan \\
$^2$ College of Science and Technology, Nihon University, Funabashi, 
         Chiba 274-8501, Japan \\
$^3$ Department of Physics, Isfahan University of Technology, 
         Isfahan 84156-83111, Iran \\
$^4$ Department of Physics, University of Tokyo, Tokyo 113-0033, Japan 
} 


~~~We have predicted few-body deeply bound kaonic states [1], starting from the Ansatz that $\Lambda (1405)$ is a $K^-p$ quasi-bound state of 1405 MeV/$c^2$ mass and 40 MeV width. Recently it was pointed out that chiral dynamics predicts the mass of $\Lambda (1405)$ to be 1420 MeV/$c^2$ or higher [2]. The position of the $K^-p$ quasi-bound state is virtually important for formation and structure of kaonic nuclear clusters. In order to solve the current debate on the $K^-p$ quasi-bound state, we have proposed to measure the $T_{21}(=T_{\Sigma \pi \leftarrow K^-p})$ $\Sigma \pi$ invariant-mass spectrum from stopped $K^-$ absorption in deuteron [3,4]. 

~~~Figure 1 compares the $\Sigma \pi$ invariant-mass spectra of our case (AMY[5]) and of Hyodo-Weise's case (H-W[2]). We have proven that a "optical" relation, which is denoted in Fig. 1, connects the $T_{21}$ invariant mass to the imaginary part of the $\bar KN$ scattering amplitude [3]: the observation of the $T_{21}$ spectrum is nothing but the observation of the $K^-p$ quasi-bound-state mass under current debate, 1405 MeV/$c^2$ or 1420 MeV/$c^2$, shown in the right panel of Fig. 1. The stopped $K^-$ on $d$ experiment [4] would provide a decisive datum for mass discrimination of the $\Lambda (1405)$, which plays an essential role in forming kaonic nuclear clusters. 

\begin{figure}[h]
\centering
  \includegraphics[height=.33\textheight]{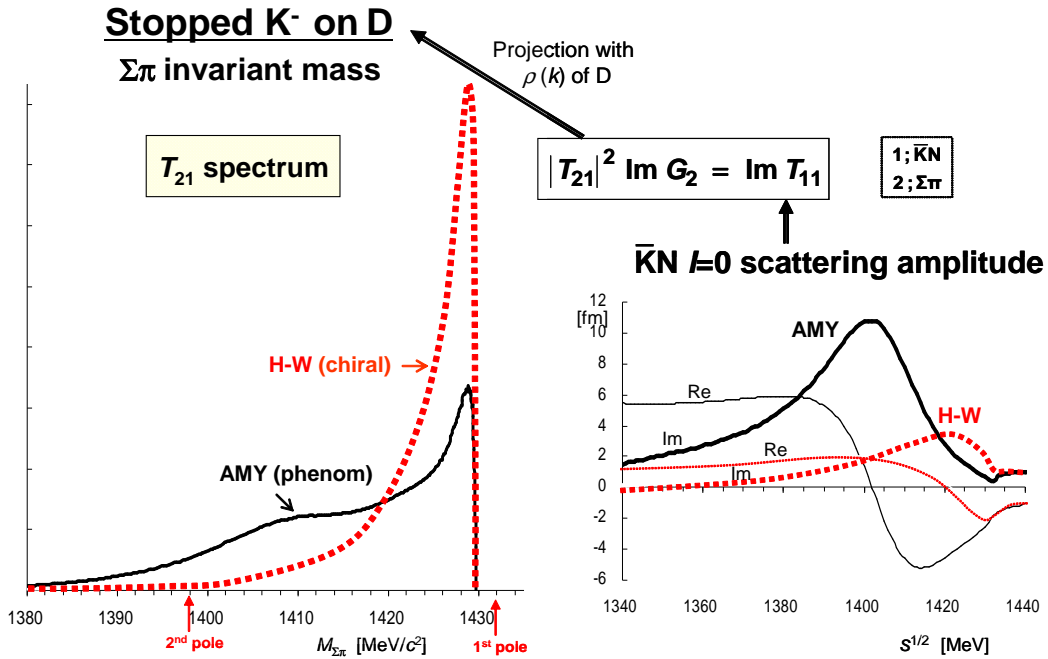}
  \caption{$\Sigma \pi$ invariant-mass spectrum from $K^-d$, and $\bar KN$ $I=0$ scattering amplitudes. }
\end{figure}

\vfill  

\noindent{\bf References }
\begin{description}
\setlength\itemsep{-3pt}

\item{[1]} Y. Akaishi and T. Yamazaki, 
                    Phys. Rev. C {\bf 65} (2002) 044005.
\item{[2]} T. Hyodo and W. Weise, Phys. Rev. C {\bf 77} (2008) 035204.
\item{[3]} J. Esmaili, Y. Akaishi and T. Yamazaki, 
                    arXiv:0909.2573 [nucl-th].
\item{[4]} T. Suzuki, J. Esmaili and Y. Akaishi, 
                    EPJ Web of Conference {\bf 3} (2010) 07014.
\item{[5]} Y. Akaishi, K.S. Myint and T. Yamazaki,
                    Proc. Jpn. Acad. B {\bf 84} (2008) 264.

\end{description}


\setcounter{equation}{0} 
\setcounter{figure}{0}
\clearpage

\addcontentsline{toc}{section}{
{\bf Strangeness in nuclear matter with $\Lambda$-hyperons
in p+C reaction at 10 GeV/c} \\
P. Aslanyan}

%

%



\titl{Strangeness in Nuclear Matter with\\ $\Lambda$-Hyperons
in p+C reaction at 10 GeV/c}

\name{
P.Aslanyan$^1$ }

\adr{Joliot-Curie 6, Dubna, p.o. 141980, Moscow region, Russia
$^1$ Joint Institute for Nuclear Research, Russia \\

}


 Now, the designed 2m propane bubble chambers(PBC) with modern power
technologies for PC and precision digital photographic methods is a
unique multi-propose, competitive capable  and higher-informative
4$\pi$ detector [2,5,6], where a beam area  is included too, for
study of exotic multi-strange events with $V^0$($\Lambda,K^0_s$ and
$\gamma$) particles, light hyper-nucleus, ($V^0, V^0$) interactions
and other correlations (P02 J-PARC LOI). The measurement errors are
equal to $\Delta\beta$ =$0.3^o$(by a azimuthal angle), $\Delta
\alpha$=$0.6^o$(a depth angle), $\Delta p/p$=2\% and $\Delta
M_{\Lambda\pi^+}/M_{\Lambda\pi^+}$ =0.6\%. First from all of
unbeatable privilege for PBC are registration of multi-vertex or
complex decay modes(with 10-50$\mu$m space resolution) and included
of the beam range too. It is very important for $\Lambda$ hyperon
physics, because more than 70\% from $\Lambda$ hyperons are emitted
over a beam range with azimuth
 $\beta$ or polar angles $< 15^0$ in p+C reaction at 10 GeV/c.

 Strange multibaryon states[1-8] with $\Lambda$- hyperon and $K^0_s$
meson subsystems has been studied by using of data from 700000
stereo photographs or $10^6$ inelastic interactions which was
obtained from expose 2-m propane bubble chamber LHEP, JINR to proton
beams at 10 GeV/c. The observed well-known resonances $\Sigma^0$,
$\Sigma^{*+}$(1385) and $K^{*+}$(892) from PDG are good tests of
this method. A number of peculiarities were found in the effective
mass spectra for $\Lambda p$ subsystem[1-8] which are crucial
important  for nuclear and particle physics in future. The study
have been continued for basic kinematic parameters  of $\Lambda$
hyperons and $\gamma$ quanta and there are registered fluctuations
($\approx$ 3$\sigma$) on distributions  by momenta and azimuthal
angles[5-8]. There are observed  enhancements   in the effective
mass distribution with total weight for reflections of  by
$\Lambda\gamma$  decay mode from  $\Xi^0$, $\Sigma^{*0}$(1385),
$\Lambda^*(1420)$ and $\Lambda^*(1560)$ hyperons[5-8].

Strange   multi-strange clusters are an exiting possibility to
explore the properties of cold dense baryonic matter[9] and
non-perturbative  QCD.   High statistics (more than 5 times ) with
above necessary conditions there will possible to establish of
obtained results. Because only with high statistics or from many
similarly old photographs without new necessary acceptance,
resolution and methods analysis there will not possible to obtain of
new information about these objects.

\vfill  

\noindent{\bf References }
\begin{description}
\setlength\itemsep{-3pt}
\item{[1]}  P.Zh.~ Aslanyan ,Elem. Part. and Atomic Nuclei,  Vol. 40, No. 4, pp. 525-557, 2009.
\item{[2]}  P.Zh.~ Aslanyan, Proc. IUTP10, Schladming, Austria, 28-6 March, 2010.
\item{[3]} P.Zh.~ Aslanyan, Proc. on  Invisible Universe Int. Conference, Blois, France, 29-3 July, 2009.
\item{[4]} P.Zh.~ Aslanyan,  "Hadronic Atoms and Kaonic Nuclei", Trento, 12 -16 October, 2009.
\item{[5]}  P.Zh.~ Aslanyan, CBM(FAIR) meeting,  Darmstadt, GSI, 12-16 April, 2010.
\item{[6]}  P.Z.~ Aslanyan, PAC-JPARC meeting,  J-PARC, Japan, 16-18 July, 2010.
\item{[7]} P.Zh.~Aslanyan,  22nd Indian Summer School, SNP, Prague, 7-11 September, 2010.
\item{[8]} P.Zh.~Aslanyan, Int. Baldin seminar  ISHEPP-XX, JINR, Dubna, 4-9 October, 2010.
\item{[9]} Y.~Akaishi and T.~Yamazaki, Phys. Rev., {\bf C65} (2002)
             044005.; Y.~Akaishi, A.~Dote and T.~Yamazaki,
             Phys. Lett., {\bf B613} (2005) 140.
\end{description}

 
\setcounter{equation}{0} 
\setcounter{figure}{0}
\clearpage

\addcontentsline{toc}{section}{
{\bf FINUDA hypernuclear spectroscopy}\\
G.~Bonomi for the FINUDA Collaboration}

%





\titl{FINUDA hypernuclear spectroscopy}

\name{
G.~Bonomi$^{1,2}$ for the FINUDA Collaboration
}

\adr{
$^1$ {Dipartimento di Ingegneria Meccanica e Industriale, University of Brescia, Brescia, Italy }\\
$^2$ {INFN Sezione di Pavia, Pavia, Italy }
\
}

The creation of an hypernucleus [1], that is a nucleus in which a nucleon is replaced by an hyperon, requires the injection of ${\it strangeness}$ into the nucleus. This is possible in different ways [2], mainly using a $\pi^+$ or a $K^-$ beam on fixed targets; recently also electron beams have been used. The FINUDA experiment at the DA$\Phi$NE $\Phi$ accelerator machine of the  INFN ``Laboratori Nazionali di Frascati'' produced $\Lambda$-hypernuclei by stopping, in thin targets (0.1-0.2 g/cm$^2$), the negative kaons originating from the $\Phi$ decay through the strangeness-exchange reaction $K^-_{stop} + ^AZ \rightarrow ^A_{\Lambda}Z + \pi^-$, where $^AZ$ indicates the target nucleus and $^A_{\Lambda}Z$ the $\Lambda$ hypernucleus in which a $\Lambda$ particle replaced a neutron. FINUDA unconventional and innovative apparatus allowed the positioning of 8 different target modules around the interaction region. In this way different targets could be studied contemporaneously, with the same apparatus and with the same analysis technique, allowing for a direct comparison between distinct materials. In particular FINUDA could study the production of $\Lambda$-hypernuclei on $^7$Li, $^9$Be, $^{12}$C, $^{13}$C and  $^{16}$O targets. Both the $\Lambda$ binding energy and the hypernuclei production probabilities have been measured. The formation probability as a function of the atomic mass number is summarized in the figure below. The new measurements on $^7_{\Lambda}$Li, $^9_{\Lambda}$Be, $^{13}_{\Lambda}$C and $^{16}_{\Lambda}$O along with previous measurements allowed  for a meaningful study of the formation of p-shell hypernuclei from the two-body capture of $K^-$ at rest giving for the first time the possibility of disentangling the effects due to atomic wave-function of the captured $K^-$ from those due to the pion optical nuclear potential and from those due to the specific hypernuclear states. The results have been submitted for publication [3].

\begin{figure}[h]
\centering
  \includegraphics[height=.27\textheight]{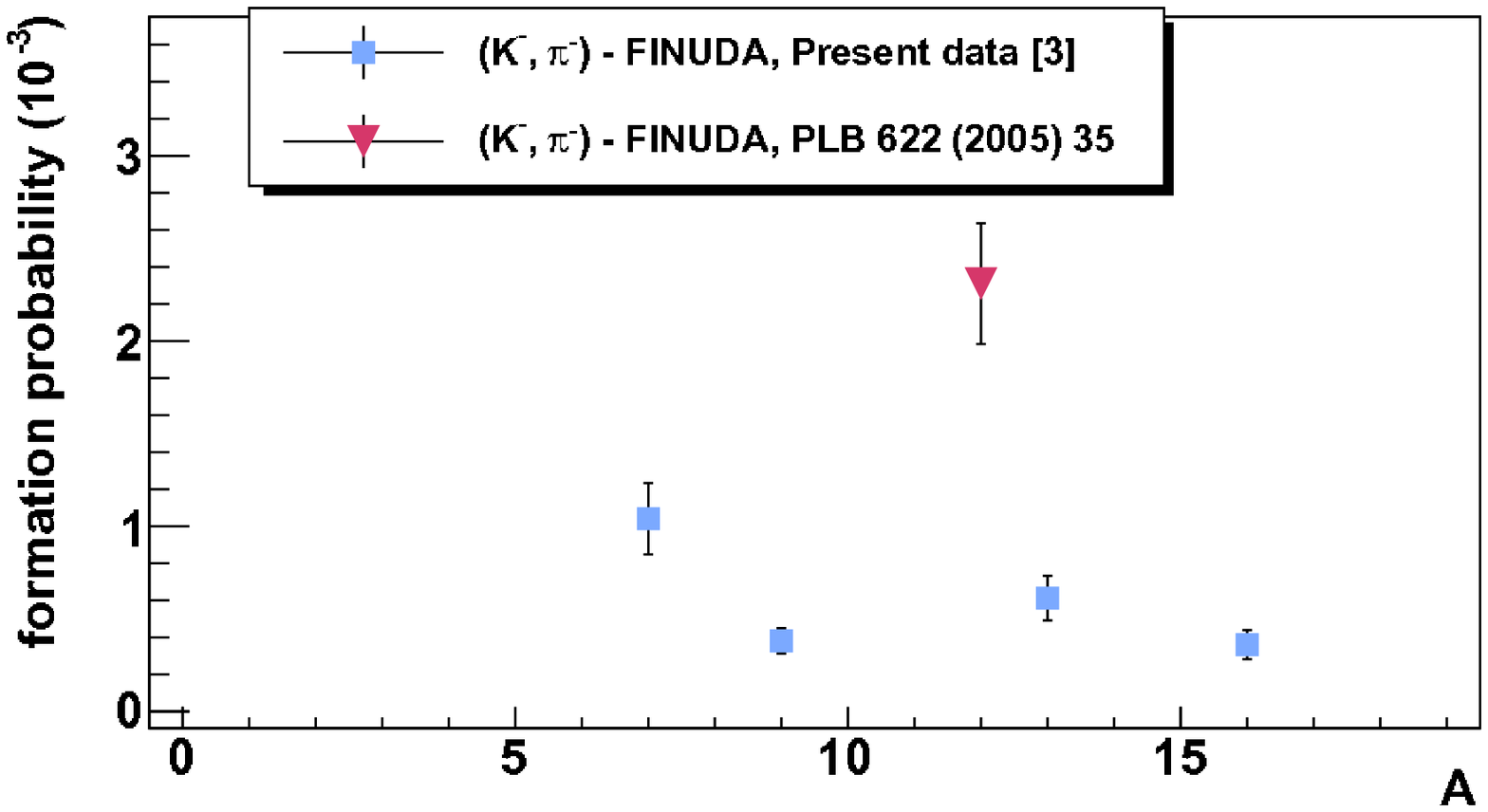}
  \caption{Hypernuclei formation probability as a function of the atomic mass number A.}
\end{figure}

\vfill  

\noindent{\bf References }
\begin{description}
\setlength\itemsep{-3pt}
\item{[1]} M. Danysz, J. Pniewski, Philos. Mag. {\bf 44} (1953) 348.
\item{[2]}
O. Hashimoto and H. Tamura, {\it Spectroscopy of $\Lambda$ hypernuclei}, Prog. Part. and Nucl. Phys. {\bf 57} (2006) 564-653.
\item{[3]} M. Agnello et al., arXiv:1011.2695v1 [nucl-ex] 11 Nov 2010.

\end{description}

 
\setcounter{equation}{0} 
\setcounter{figure}{0}
\clearpage

\addcontentsline{toc}{section}{
{\bf Strangeness production by antiproton annihilation at
rest in  H, D, $^3He$ and $^4He$}\\
T.~Bressani}

%





\titl{STRANGENESS PRODUCTION BY ANTIPROTON ANNIHILATION AT REST IN H, D, $^3He$ AND $^4He$}

\name{
T.~Bressani$^{1,2}$
}

\adr{
$^1$ Dipartimento di Fisica Sperimentale dell' Universit\`a, Torino (Italy)\\
$^2$INFN Sezione di Torino, Via P. Giuria 1, Torino (Italy) 
}


The dependence of the K$^+$ and K$^-$ production on the number of nucleons involved in the annihilation process is investigated experimentally in the $\bar{p}$ annihilation at rest on hydrogen, deuterium, $^3$He and $^4$He gas targets [1].
The data (about 3 Millions of annihilation events) were collected at the LEAR machine at CERN by using the state--of--art detector OBELIX. Annihilations with any number of prongs (charged pions and kaons, protons and deuterons) are analyzed. Events with and without emission of fast neutrons (that is neutrons involved in the annihilation process) are recognized. The results put in evidence that the strangeness production is lower or higher depending on the reaction channel. As a general trend, the strangeness production is higher in events without neutral mesons and still higher in events with the involvement of a higher number of nucleons. Both K$^+$ and K$^-$ productions increase with the number of involved nucleons, but K$^+$ much more. The maximum K$^+$ production is observed in the reaction K$^+$2$\pi^+$2$\pi^-3$n on $^4$He (with the involvement of 3-4 nucleons); compared with the production on hydrogen K$^+$$\pi^+$2$\pi^-$, the production on $^4$He is higher by a factor of 31.7$\pm$5.5. Following some theoretical speculations this enhancement factor is too high to be explainable in terms of hadronic interactions and could be interpreted as a signature of quark deconfinement and of formation of a quark-gluon plasma [2]. The pion spectra, measured also in coincidence with a p, a K$^{\pm}$ and a pK$^{\pm}$, could be fitted by a statistical model with T=140 MeV, a value close to that deduced from collisions of relativistic heavy ions.
The hypothesis of a quark-gluon plasma formation could also explain the measured production rate for a possible candidate of a K$^-$pp Antikaonic Nuclear Cluster previously observed [3] and of S=-2 Strangeness production [4].

\vfill  

\noindent{\bf References }
\begin{description}
\setlength\itemsep{-3pt}
\item{[1]} G. Bendiscioli {\it et al.}, Nucl. Phys A {\bf 815} (2009) 67.
\item{[2]} J. Rafelski and B Muller, Phys. Rev. Lett. {\bf 48} (1982) 1066.
\item{[3]} G. Bendiscioli {\it et al.}, Nucl. Phys. A {\bf 789} (2007) 222.
\item{[4]} G. Bendiscioli {\it et al.}, Nucl. Phys. A {\bf 797} (2007) 109.

\end{description}

 
\setcounter{equation}{0} 
\setcounter{figure}{0}
\clearpage

\addcontentsline{toc}{section}{
{\bf Weak decay of $\Lambda$--hypernuclei}\\
S.~Bufalino for the FINUDA Collaboration}

%





\titl{Weak decay of $\Lambda$--hypernuclei}

\name{
S.~Bufalino$^1$ for the FINUDA Collaboration
}

\adr{
$^1$ INFN Sezione di Torino, via P. Giuria 1, Torino, Italy \\
}


The information coming from the study of the $\Lambda$-hypernuclei weak decay channels complements the 
knowledge of strange nuclear 
systems obtained by both missing mass and $\gamma$-ray spectroscopy measurements. 
$\Lambda$-hypernuclei decay through both the mesonic weak decay (MWD) processes: 
\begin{eqnarray} 
^{A}_{\Lambda}\mathrm{Z} & \rightarrow & ^{A}(\mathrm{Z+1}) + \pi^{-} \ \ \ \ (\Gamma_{\pi^{-}}) \\ 
^{A}_{\Lambda}\mathrm{Z} & \rightarrow & ^{A}\mathrm{Z} + \pi^{0}  \ \ \ \ (\Gamma_{\pi^{0}})
\label{mwd} 
\end{eqnarray} 
and the non-mesonic weak decay (NMWD) processes: 
\begin{eqnarray} 
^{A}_{\Lambda}\mathrm{Z} & \rightarrow & ^{A-2}(\mathrm{Z-1}) + p + n  \ \ \ \ (\Gamma_{p}) \\ 
^{A}_{\Lambda}\mathrm{Z} & \rightarrow & ^{A-2}\mathrm{Z} + n + n  \ \ \ \ (\Gamma_{n}) \\
^{A}_{\Lambda}\mathrm{Z} & \rightarrow & ^{A-3}\mathrm{Z'} + N + N + N  \ \ \ \ (\Gamma_{2})
\label{nmwd} 
\end{eqnarray}
The channel (5) is indicated as 
two-nucleon induced (2N) decay and refers to the interaction of the $\Lambda$ with a couple of strongly 
correlated nucleons; $Z'$ stands for $Z$, $Z-1$ or $Z-2$ depending on the particular nucleons combination.  
The FINUDA experiment performed a complete analysis of the charged particles ($\pi^{-}$ and $p$) spectra 
following the MWD and NMWD of ${\mathrm{^{5}_{\Lambda}He}}$, ${\mathrm{^{7}_{\Lambda}Li}}$, ${\mathrm{^{9}_{\Lambda}Be}}$, $
{\mathrm{^{11}_{\Lambda}B}}$, ${\mathrm{^{12}_{\Lambda}C}}$, ${\mathrm{^{13}_{\Lambda}C}}$, ${\mathrm{^{15}_{\Lambda}N}}$ and  $
{\mathrm{^{16}_{\Lambda}O}}$ hypernuclei. \\ 
MWD spectra and decay rates have been obtained for  $^{7}_{\Lambda}\mathrm{Li}$, $^{9}_{\Lambda}\mathrm{Be}$, $^{11}_{\Lambda}\mathrm{B}$ and $^{15}_{\Lambda}\mathrm{N}$ for the first time and compared with 
previous measurements and calculations. The spin-parity assignment $J^{\pi}(^{15}_{\Lambda}\mathrm{N}_{g.s.}) = {3/2}^+$ 
was made for the first time and the results have been published [1].\\
The FINUDA Collaboration also analyzed the proton energy spectra of ${\mathrm{^{5}_{\Lambda}He}}$, ${\mathrm{^{7}_{\Lambda}Li}}$, ${\mathrm{^{9}_{\Lambda}Be}}$, ${\mathrm{^{11}_{\ \Lambda}B}}$,
 ${\mathrm{^{12}_{\ \Lambda}C}}$, ${\mathrm{^{13}_{\ \Lambda}C}}$, ${\mathrm{^{15}_{\ \Lambda}N}}$ and ${\mathrm{^{16}_{\ \Lambda}O}}$ with good resolution ($\Delta$p/p=2\% FWHM for protons of 80 MeV) and with a detection threshold of 15 MeV. All measured spectra showed a similar behaviour, i.e. a bump at about 80 MeV, roughly at the energy expected from reaction (3)(with an incertitude of about 10 MeV due to several nuclear structure and interaction effects). The bump is quite well defined in the high energy portion, whereas at low energies it is blurred in a continuum generated by FSI, superimposed to the 2N-induced NMWD contribution. With very simple hypotheses and a model--independent method the contributions from FSI and 2N-induced NMWD were disentangled,  providing a value of  $\Gamma_{2N}$/$\Gamma_{p}$ of 0.43$\pm$0.25 and $\Gamma_{2N}$/$\Gamma_{NMWD}$ of 0.24$\pm$0.10 [2]. 
This method was recently improved with the further detection of a neutron, from which we determined the value of $\Gamma_{2N}$/$\Gamma_{NMWD}$ with an error reduced of a factor about two respect to the previous FINUDA determination [2].
In fact, in spite of the low statistics, we determined the value of  $\Gamma_{np}$/$\Gamma_{p}$=0.33$\pm$${0.07_{stat}}^{+0.04_{sys}}_{{-{0.03_{sys}}}}$ and of $\Gamma_{np}$/$\Gamma_{NMWD}$=0.23$\pm$${0.04_{stat}}^{+0.03_{sys}}_{{-{0.01_{sys}}}}$. We also extracted $\Gamma_{2N}$/$\Gamma_{NMWD}$=0.28$\pm$${0.05_{stat}}^{+0.04_{sys}}_{{-{0.01_{sys}}}}$. The value is in agreement, within the errors, with previous evaluations, model dependent or not, and with theoretical calculations. 

\vfill  

\noindent{\bf References }
\begin{description}
\setlength\itemsep{-3pt}
\item{[1]} M. Agnello {\it et al.}, Phys. Lett. B {\bf 681} (2009) 139.
\item{[2]}  M. Agnello {\it et al.}, Phys. Lett. B {\bf 685} (2010) 247.
\end{description}

 
\setcounter{equation}{0} 
\setcounter{figure}{0}
\clearpage

\addcontentsline{toc}{section}{
{\bf $\Lambda$ hypernuclear production revisited}\\
A.~Ciepl\'{y}, A.~Gal, V.~Krej\v{c}i\v{r}\'{\i}k}

%





\titl{$\Lambda$ hypernuclear production revisited
\footnote{The work was supported by Grant Agency of the Czech Republic, grant 202/09/1441.}
}

\name{
A.~Ciepl\'{y}$^{1}$, A.~Gal$^{2}$, V.~Krej\v{c}i\v{r}\'{\i}k$^{1}$
}

\adr{
$^1$ Nuclear Physics Institute, 25068 \v{R}e\v{z}, Czech Republic \\
$^2$ Racah Institute of Physics, The Hebrew University, 91904 Jerusalem, Israel
}


The new experimental data [1] on the $\Lambda$-hypernuclear production 
in ($K^{-}_{\rm stop}$,$\pi^{-}$) reactions allow to study the $A$-dependence of the formation 
rates for the $p$-shell nuclear targets. Within a framework of the distorted wave impulse approximation 
the nuclear capture rate per stopped kaon $R_{fi}/K$ can be expressed [2] as a product of three terms,
a kinematical factor, the elementary branching ratio (BR) of the process $R(K^-N\rightarrow\pi\Lambda)$, 
and a rate per hyperon $R_{fi}/Y$. Assuming a constant BR generated by an effective chiral model,  
taken at the $\bar{K}N$ threshold and for an intermediate nuclear density, the $A$-dependence 
of the formation rates is solely driven by the last term, the rate per hyperon, 
that contains the overlap of the initial and the final state wave functions. In Fig.~1 we 
show the calculated formation rates (per stopped kaon) in comparison with the FINUDA data for 
nuclear targets from $^{7}$Li to $^{16}$O. The presented theoretical rates were obtained [3]
for a fixed pion potential and for two different kaon-nuclear optical potentials, a deep 
density dependent phenomenological potential fitted to kaonic atoms data and a shallow 
chirally motivated potential.

\vspace*{-8mm}
\begin{figure}[h]
\centering
\includegraphics[height=.24\textheight]{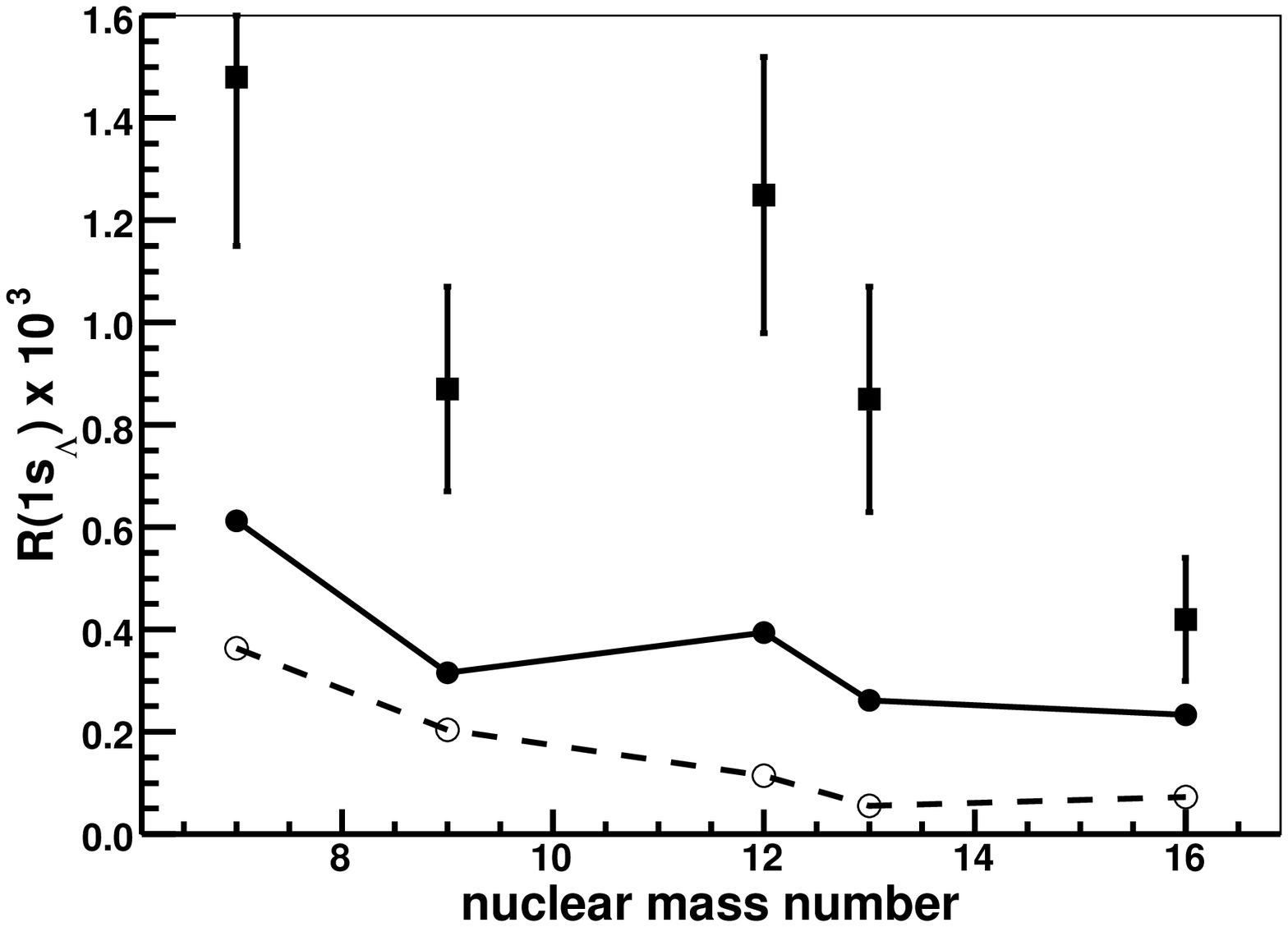}
\vspace*{-4mm}
\caption{\small $1s_{\Lambda}$ formation rates in $K^-$ capture at rest on $p$ shell 
nuclei. The experimental rates derived from FINUDA data [1] are compared with 
our results [3] as calculated for a shallow $K^-$ nuclear potential (solid line) and 
for a deep potential (dashed line).}
\end{figure}

\vspace*{-2mm}
The calculated capture rates generally underestimate the measured ones, 
the deeper the $K^-$ potential, the smaller is the capture rate. For a given $K^-$ 
optical potential the rates decrease as a function of $A$, with the fractional difference 
between the rates calculated for the two extreme $K^-$ optical potentials, 
the shallow chiral one and the deep phenomenological one, growing up steadily with $A$.
We note that a final conclusion may be reached only when the energy and density dependence of the
BR below threshold is considered within a theoretical model [4].

\vfill  

\noindent{\bf References }
\vspace*{-2mm}
\begin{description}
\setlength\itemsep{-3pt}
\item{[1]} M.~Agnello {\it et al.} (FINUDA Collaboration), ePrint arXiv:1011.2695 [nucl-ex]; 
M.~Agnello {\it et al.}, Phys.~Lett.~B {\bf 622} (2005) 35.
\item{[2]} A.~Gal and L.~Klieb, Phys.~Rev.~C {\bf 34} (1986) 956.
\item{[3]} V.~Krej\v{c}i\v{r}\'{i}k, A.~Ciepl\'{y} and A.~Gal, Phys.~Rev.~C {\bf 82} (2010) 024609.
\item{[4]} A. Ciepl\'{y}, E. Friedman, A. Gal, D. Gazda and J. Mare\v{s}, work in
preparation.
\end{description}

\setcounter{equation}{0} 
\setcounter{figure}{0}
\clearpage

\addcontentsline{toc}{section}{
{\bf The SIDDHARTA2 experiment}\\
C.~Curceanu  on behalf of the SIDDHARTA2 Collaboration}


%





\titl{The SIDDHARTA2 experiment}

\name{
C.~Curceanu$^{1}$  on behalf of the SIDDHARTA2 Collaboration
} 

\adr{
$^1$  INFN, Laboratori Nazionali di Frascati, CP 13,
 Via E. Fermi 40, I-00044, Frascati (Roma), Italy \\

}


The SIDDHARTA experiment was successfully performing kaonic helium (3 and 4) 
and kaonic hydrogen measurements on the DA$\Phi$NE Collider at LNF-INFN in
2009 (see contributions of T. Ishiwatari, D. Sirghi and S. Okada),
 proving that the used technique, namely triggered Silicon Drift Detectors
 combined with the kaon beam delivered at DA$\Phi$NE is an excellent opportunity for these type of measurements.
Based on this success, the proposal for a substantial upgrade of the SIDDHARTA apparatus, becoming SIDDHARTA2, was put forward [1] by an enriched collaboration. 
The upgrade is concerning:
\begin{itemize}
\item  enhancement of the signal, by using a target allowing to stop more kaons inside (higher density and higher solid angle)
\item  background reduction (by reinforced shielding and optimized trigger system)
\item possibility to use solid targets  with use of new detectors for
  higher energy photons
\end{itemize}
With this setup a new series of measurements could be then performed:
\begin{itemize}
\item kaonic deuterium transitions to the 1 s level, which, 
taken together with the kaonic hydrogen results obtained by SIDDHARTA will provide the isospin-dependent antikaon-nucleon scattering lengths, fundamental information for the study of the low-energy QCD in the strangeness sector
\item other type of kaonic atoms measurements: going from light atoms to heavier ones, in order to perform more precise measurements than the existing ones (performed 20-30 years ago), so to provide theoreticians with a consistent data base
\item feasibility studies for sigmonic atoms
\item feasibility study for the precision measurement of the charged kaon mass
\end{itemize}
The aim of the SIDDHARTA2 Collaboration is to prepare the upgraded setup
within the end of 2011, such as to be ready to install and start data taking
on DA$\Phi$NE in 2012.

\vfill  

\noindent{\bf References }
\begin{description}
\setlength\itemsep{-3pt}
\item{[1]} The upgrade of the SIDDHARTA apparatus for an enriched scientific case. 14 June 2010
Presented to the LNF International Scientific Committee, 24 June 2010
\end{description}

 
\setcounter{equation}{0} 
\setcounter{figure}{0}
\clearpage

\addcontentsline{toc}{section}{
{\bf Updated results for $\Lambda(1405) \rightarrow \Sigma^{0}\pi^{0}$ with HADES}\\
E. Epple  for the HADES collaboration}

%




\titl{Updated results for \pmb{$\Lambda(1405) \rightarrow \Sigma^{0}\pi^{0}$} with HADES}
\name{Eliane Epple for the HADES Collaboration}
\adr{Excellence Cluster ``Universe", TU M\"unchen, Boltzmannstr. 2, 85748 Garching, Germany
}
The decay of the $\Lambda(1405)$ hyperon into $\Sigma^{0} \pi^{0}$ is of special interest as it probes the 
\textbf{\textit{I}}$=0$ component exclusively and thus allows the extraction of a pure $\Lambda(1405)$ line shape [1].
The measurement of this decay has so far only been performed with low statistic [2].\\
These facts give the motivation for investigating the particular decay of
the $\Lambda(1405)$ into $\Sigma^{0}\pi^{0}$ in p+p collisions at $E_{kin}$=3.5 GeV, measured by the $\textbf{H}$igh $\textbf{A}$cceptance $\textbf{D}$i-$\textbf{E}$lectron $\textbf{S}$pectrometer ($\textbf{HADES}$).
Most challenging in the data analysis is to enhance the
final signal of the reaction (1) and suppress the reaction (2).
\begin{flalign}
p+p \rightarrow \Lambda(1405)&+p+K^{+}\rightarrow(p,\pi^{-},\pi^{0},\gamma)+p+K^{+},  \\
p+p \rightarrow \Sigma(1385)^{0}&+p+K^{+}\rightarrow(p,\pi^{-},\pi^{0})+p+K^{+}.
\end{flalign}
This is only possible by a cut on the observable of the missing mass of $p,K^{+},p,\pi^{-}$ ($\Delta M_{p,K^{+},p,\pi^{-}}$).
Due to the finite mass resolution of the spectrometer it is not possible to cut away all signal from the production of $\Sigma(1385)^{0}$.
Out of this reason and the fact that several other reactions and also $K^{+}$ misidentification contribute to
the final $\Lambda(1405)$ spectrum (see fig. 1), we exploited simulations to evaluate the line shape of these contributing channels. 
Thus, simulations of the following reactions have been performed: $p+p$ $\rightarrow$\\
1) $\Lambda K^+p$, 2) 
$\Lambda(1405)K^+p$, 3) $\Sigma^0K^+p$, 4) $\Lambda K^+p \pi^0$, 5) $\Sigma^+K^+p\pi^-$, 6) $\Sigma^+K^+p\pi^-\pi^0$, 7) 
$\Sigma^0K^+p\pi^-\pi^+$, 8) $\Sigma(1385)^{0}K^{+}p$, 9) $\Lambda K^+p\pi^-\pi^+$, 10) $\Lambda(1520)pK^+$, 
11) $\Sigma^0K^+p\pi^0$. 
We have fitted all the simulated contributions as well as the misidentification background to experimental spectra of
$\Delta M_{p,K^{+},p,\pi^{-}}$ to extract the strengths of 1)-11). The result of the fit is shown in fig. 2.
For more details see [3]. With the modeling of the background strength it is possible to subtract it from the final spectrum to obtain a pure $\Lambda(1405)$ signal.
\begin{figure}[h]
\begin{center}
\begin{minipage}[t]{0.46\textwidth}
\center
\includegraphics[width=6.34 cm]{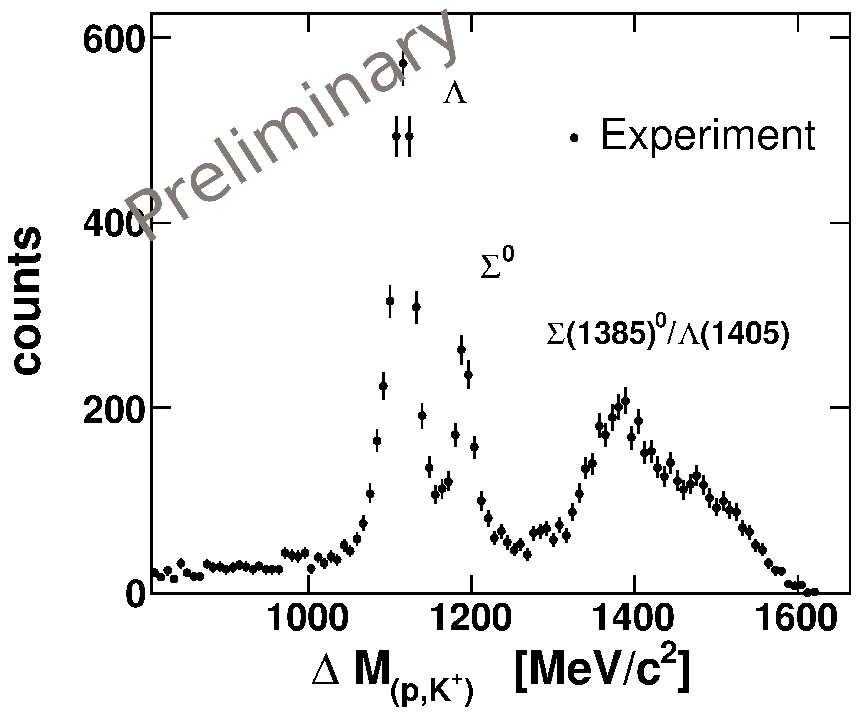}
\caption{\label{figMMpK} Missing mass of $p,K^{+}$ showing several channels which fulfill the event selection constraints.}
\end{minipage} 
\hspace{0.2 cm}
\begin{minipage}[t]{0.46\textwidth}
\center
\includegraphics[width=6.34 cm]{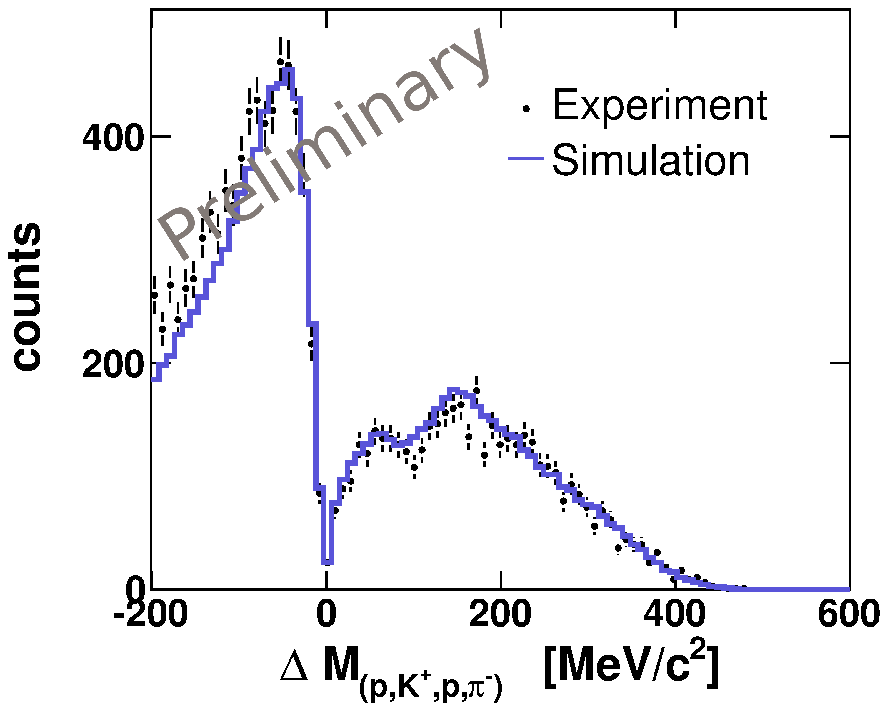}
\caption{\label{KA} Missing mass of all four charged particles $\Delta M_{(p,K^{+},p,\pi^{-})}$ together with the result of the fit.}
\end{minipage}
\end{center}
\end{figure}
\vfill  
\noindent{\bf References }
\begin{description}
\setlength\itemsep{-3pt}
\item{[1]} J. C. Nacher, E. Oset, H. Toki and A. Ramos,  Phys. Lett. {\bf B455} (1999) 55-61.
\item{[2]} I. Zychor {\it et al.}, Phys. Lett. {\bf B660} (2008) 167-171.
\item{[3]} E. Epple {\it et al.}, [HADES Collaboration], Int. J. Mod. Phys. {\bf A26} (2011) 616 .
\end{description}

 
\setcounter{equation}{0} 
\setcounter{figure}{0}
\clearpage

\addcontentsline{toc}{section}{
{\bf Low energy kaon scattering, an experimental overview}\\
A. Filippi}

%





\titl{Low Energy Kaon Scattering, an experimental overview}

\name{
A. Filippi$^{1}$
}

\adr{
$^1$ I.N.F.N. Torino, via P. Giuria, 1, 10125 Torino, Italy \\
}


The present knowledge of total and differential cross sections of reactions induced by low energy kaons
is scarce, since the available experimental data are very few. Below 150 MeV/$c$, in fact, only
dated measurements of $K^\pm p$ total and elastic cross sections exist, and they bear large errors.
Nonetheless, precise cross section measurements at low momenta (especially very close
to threshold) are needed as inputs for theoretical models of $\overline K (K) N$
interactions [1]. 

The DA$\Phi$NE $\phi$ factory in LNF is a unique place where studies of low energy kaon
scattering could be pursued. In DA$\Phi$NE, $\phi$'s are
produced with a small boost (12 MeV/$c$), so that the charged kaons from their decay have a momentum
in the range $(115-140)$ MeV/$c$. Low momentum kaon interactions could therefore be studied;
measurements of particles with such low momenta represent, clearly, a big experimental challenge.

A project for a detector able to accomplish this task has been proposed
(IKON, Interacting Kaons On Nucleons).
Its basic layout is a low-mass multilayer modular Si-vertex detector 
with cylindrical geometry, hosting a gaseous Hydrogen or
Deuterium target (at 2-3 bar pressure). Two inner layers provide the incoming
kaon tracking. Beyond the target, outer layers can track the final products of the reaction.
The dimensions of the full detector are such as
to allow to nest it in the hollow space (with a properly modified beam pipe)
at the center of the KLOE apparatus, in a similar way as proposed by AMADEUS to pursue
their primary stopped-kaon physics program.
Occupancy studies show that a total
number of 100 200 $\mu$m$\times (6\times 5)$ cm$^2$ Si-modules, arranged in coaxial prisms, would
be enough to cover the 75\% of the $K^\pm$ beams acceptance. The Si-modules have a spatial resolution
($<50\ \mu$m) such as to allow a precise  
 secondary vertex reconstruction and are able to provide charged particle 
 identification by means of $dE/dx$ measurements. 
The use of self-triggering Si-modules would moreover make an additional start-detector unnecessary.

Complete measurements of several kaon induced
reactions could be performed at the same time, exploiting the KLOE calorimeter for the
neutral particles detection. One could for instance address to:
\begin{itemize}
  \item the elastic $K^\pm p\rightarrow K^\pm p$ reaction:  an integrated luminosity of
3 fb$^{-1}$ would be enough to perform total cross section measurements in 
  at least two (possibly three) momentum bins, with a percent precision, and of 
  differential cross section with a precision at the level of, respectively, 
  7\% for $K^+p$ and 5\% for $K^-p$;
  \item the reaction on deuteron $K^-d\rightarrow \Sigma^{\pm,\mp,0}\pi^{\mp,\pm,0} n$ at the
  $\Lambda(1405)$ peak, with the detection of forward neutrons only, which would allow to
  disentangle the
  $\Lambda(1405)$ peak from the Quasi-Free continuum [2]. In a 4 fb$^{-1}$ data taking
  about 500 events could be collected, in both neutral
  and charged $\Lambda(1405)$ decay channels.
\end{itemize}

 Such a modular vertex detector could be fruitfully merged with the proposed AMADEUS set-up to perform 
 both the stopped- and in-flight kaon interaction studies.

\vfill  

\noindent{\bf References }
\begin{description}
\setlength\itemsep{-3pt}
\item{[1]} B. Borasoy, U.G. Meissner, R. Nissler, {\it Phys. Rev.} {\bf C64} (2006), 055201
\item{[2]} D. Jido {\it et al.}, arXiv:1008.4423v2 [nucl-th]
\end{description}

 
\setcounter{equation}{0} 
\setcounter{figure}{0}
\clearpage

\addcontentsline{toc}{section}{
{\bf Unexpected features of $K^\pm$ interaction with nuclei at low energies}\\
E.~Friedman}




%






























\titl{Unexpected features of $K^\pm$ interaction with nuclei at low energies}




\name{





E.~Friedman$^1$

}




\adr{

$^1$ Racah Institute of Physics, the Hebrew University, Jerusalem, Israel \\

}




We discuss a conflict between phenomenological potentials obtained
from global fits to strong-interaction observables in kaonic atoms 
and potentials constructed from more fundamental approaches to the
$\bar K N$ interaction.  Analyses of large data sets encompassing
the whole of the periodic table use optical potentials which 
could reveal characteristic features of the 
interaction [1,2]. With a $t\rho$ approach to the optical potential
and allowing a density-dependent effective amplitude $t(\rho$) [3],
greatly improved fits to the data are possible, compared to a fixed-$t$
approach. However, the depths of the real potential are typically four
times larger than what chiral models predict [2,4]. Alternatively,
with $t(\rho)$ predicted by such models the agreement between
calculations and kaonic atom experiments is rather poor.

Integral quantities which usually characterize optical potentials are
the volume integral per target nucleon and the {\it rms} radius 
of the potential, separately for the real and for the imaginary parts.
Ref.[3] showed that the potentials may be classified by a 
`compression' or `inflation' relative to the corresponding {\it rms}
radius of the target nucleus, with the family with
compression of the real potential producing the
best fit to the data (and having the deepest real potential). 
Attention has been focused mainly on the depth of the real potential, 
because of possible consequences for kaon condensation in neutron stars
and more recently because of speculations on the possible
existence of $K^-$-nuclear strongly-bound clusters [5]. 
However, the systematics of {\it rms}
 radii could be the more
significant result.
Obviously inflation or an increased radius 
is expected for a potential that is obtained by folding a $K^-N$
interaction with the nuclear density. 
Additionally the radius of the potential increases
when the interaction strength decreases with increased
density.
In contrast,
 compression is expected when the
underlying interaction strength increases with  density so as
to offset the inflation due to the finite-range folding. 
 These features are  indeed observed in amplitudes predicted 
by chiral models [4], but the failure to fit experimental results is
still an open problem, as is the question of the depth of the 
real potential.

Finally we mention an open problem with the $K^+$-nucleus interaction
between 400 and 700 MeV/c where some 20\% of extra reactivity is still
not accounted for [2].






\vfill  





\noindent{\bf References }

\begin{description}

\setlength\itemsep{-3pt}


\item{[1]} C.J.~Batty {\it et al.}, Phys. Reports {\bf 287} (1997) 385.

\item{[2]} E.~Friedman, A. Gal,  Phys. Reports {\bf 452} (2007) 89.

\item{[3]} E.~Friedman {\it et al.}, Phys. Lett. B {\bf 308} (1993) 6;
Nucl. Phys. A {\bf 579} (1994) 518.

\item{[4]} A.~Ciepl\'y {\it et al.}, Nucl. Phys. A {\bf 696} (2001) 173.

\item{[5]} Y.~Akaishi, T.~Yamazaki,  Phys. Rev. C {\bf 65},
044005 (2002).

\end{description}


\setcounter{equation}{0} 
\setcounter{figure}{0}
\clearpage

\addcontentsline{toc}{section}{
{\bf Strangeness --1 dibaryons revisited}\\
A. Gal} 




\titl{Strangeness --1 dibaryons revisited} 

\name{Avraham Gal$^1$} 

\adr{$^1$ Racah Institute of Physics, the Hebrew University, Jerusalem 91904, 
Israel \\} 

Following a brief overview of dibaryon theoretical expectations, the notion of 
pion assisted dibaryons, $\pi BB^\prime$, is introduced [1]. The idea is to 
enhance the binding of a $L=0$ $BB^\prime$ configuration through the strong 
attraction provided by $p$-wave $\pi B$ resonances. A possible quasibound 
$\pi\Lambda N$ stretched configuration $(I=\frac{3}{2},J^P=2^+)$ has been 
recently studied [2,3] by solving three-body Faddeev equations with 
$^3S_1-{^3D}_1,~\Lambda N -\Sigma N$ coupled channels chirally based local 
interactions, and coupled $\pi Y$ ($Y\equiv\Lambda,\Sigma$) and $\pi N$ 
separable $p$-wave interactions fitted to the position and decay parameters 
of the $\Sigma(1385)$ and $\Delta(1232)$ resonances, respectively. 
The results exhibit strong sensitivity to the $p$-wave $\pi Y$ interaction, 
the least phenomenologically constrained interaction in this calculation, 
with a $\pi\Lambda N$ quasibound state (denoted $\cal D$) persisting over 
a wide range of acceptable parametrizations. $\cal D$ might prove the lowest 
lying ${\cal S}=-1$ dibaryon, well below the anticipated $(I=\frac{1}{2},
J^P=0^-)$ quasibound $K^-pp$.   
\newline 
\newline 
The dibaryon $\cal D$ decays to a $d$-wave $I=\frac{3}{2}$ $\Sigma N$ 
scattering state and perhaps also to $(\pi \Lambda N)_{I=\frac{3}{2}}$ 
if it corresponds to $\pi \Sigma N$ quasibound state above the $\pi \Lambda N$ 
threshold. To search for $\cal D$, the following reactions are possible: 
\begin{equation} 
K^- + d \to  {\cal D}^- + \pi^+~, \,\,\,\,\,\,\,\,\,\, 
\pi^- + d \to {\cal D}^- + K^+~,  
\label{eqKpiK}
\end{equation}   
\begin{equation} 
p + p \to {\cal D}^+ + K^+~.  
\label{eqpp} 
\end{equation} 
\newline 
\newline 
Extensions from ${\cal S}=-1$ are possible to the ${\cal C} \neq 0$ charm 
sector. One could also explore baryon resonances of a similar structure, 
such as a pion assisted $(I=1,J^P={\frac{3}{2}}^-)$ ${\overline K} N \pi$ 
quasibound state around 1540 MeV [4].

\vfill  

\noindent{\bf References} 
\begin{description} 
\setlength\itemsep{-3pt} 
\item{[1]} A.~Gal, arXiv:1011.6322 (nucl-th). 
\item{[2]} A.~Gal and H.~Garcilazo, Phys. Rev. D {\bf 78} (2008) 014013. 
\item{[3]} H.~Garcilazo and A.~Gal, Phys. Rev. C {\bf 81} (2010) 055205. 
\item{[4]} A.~Gal and H.~Garcilazo, in preparation. 
\end{description} 


\setcounter{equation}{0} 
\setcounter{figure}{0}
\clearpage

\addcontentsline{toc}{section}{
{\bf Pseudopotential approach for $\bar{K}$-nuclei based on Chiral $\bar{K}N$ amplitude}\\
D.~Gazda}

%





\titl{Pseudopotential Approach for $\bar{K}$-nuclei Based on Chiral $\bar{K}N$ Amplitude}

\name{
D.~Gazda $^{1,2}$
}

\adr{
$^1$ Nuclear Physics Institute, 25068 \v{R}e\v{z}, Czech Republic \\
$^2$ Faculty of Nuclear Sciences and Physical Engineering, Czech Technical University in Prague, 11519 Prague, Czech Republic
}


In this contribution I report on our recent study of $\bar{K}$-nuclear
states. The model adopted in our calculations is the pseudopotential approach, where
the optical potential is taken in the $t\rho$ form and introduced
selfconsistently into the Klein-Gordon equation for the $K^-$ meson. Due to the
strongly attractive nature of the $\bar{K}N$ interaction the $\bar{K}N$
scattering amplitude $t=t(\omega,\rho)$ is calculated nonperturbatively within
the framework of multichannel scattering constrained by chiral symmetry [1]. The
realistic density distribution $\rho(r)$ is taken from the RMF nuclear
phenomenology for several core nuclei across the periodic table [2]. The focal
question of this work is to explore the energy and density dependence of the
in-medium chiral $\bar{K}N$ amplitude and its implications for $K^-$-nuclear
bound states.
In Table \ref{tab:res} we present selected results for the binding energies
$B_{K^-}$ and conversion widths $\Gamma_{K^-}$ of $1s$ $K^-$-nuclear states in
several nuclei. The energy dependence of the $\bar{K}N$ scattering amplitude
results in deep binding (80-110~MeV) and moderate absorption widths (10-30~MeV)
of the $K^-$ mesons in the nuclear medium when extrapolated to subthreshold
region. It is to be stressed that the results are calculated in the absence of
the nonmesonic conversion modes on two nucleons $\bar{K}NN\rightarrow YN$. Based
on our previous results [2], we can estimate the contribution of $2N$ absorption
modes to the $K^-$ decay width to be of the order of $\approx 30$~MeV.
Nevertheless, the two nucleon absorption processes as well as the many body
effects beyond two body $\bar{K}N$ dynamics in the nuclear medium require
further studies, as indicated by our calculations of kaonic atoms.
\renewcommand{\arraystretch}{1.15}
\begin{table}[h]
\centering
\caption{The $1s$ $K^-$ binding energies $B_{K^-}$ and conversion widths
$\Gamma_{K^-}$ for selected $K^-$-nuclei, calculated using the CS30 chiral model
of Ref.\ [1].}
\label{tab:res}
\begin{tabular}{|c|c|c|}
\hline
 & $B_{K^-}$ (MeV) & $\Gamma_{K^-}$ (MeV) \\ \hline
$^{12}$C   & 81.7  & 28.5   \\ \hline
$^{16}$O   & 76.3  & 29.6   \\ \hline
$^{40}$Ca  & 96.5  & 16.6   \\ \hline
$^{90}$Zr  & 104.3  & 10.2   \\ \hline
$^{208}$Pb & 109.0  & 9.9    \\ \hline
\end{tabular}
\end{table}

\bigskip
I would like to acknowledge and thank my collaborators A.\ Ciepl\'{y}, E.\
Friedman, A.\ Gal, and J.\ Mare\v{s}. This work was supported in part by the CTU
grant SGS10/087/OHK4/1T/14 and GACR grant 202/09/1441.

\vfill  

\noindent{\bf References }
\begin{description}
\setlength\itemsep{-3pt}
\item{[1]} A.\ Ciepl\'{y}, J.\ Smejkal, Eur.\ Phys.\ J.\ A {\bf 43} (2010) 191.
\item{[2]} D.\ Gazda, E.\ Friedman, A.\ Gal, J.\ Mare\v{s}, Phys.\ Rev.\ C {\bf 76} (2007) 055204.
\end{description}


\setcounter{equation}{0} 
\setcounter{figure}{0}
\clearpage

\addcontentsline{toc}{section}{
{\bf Five- and four-body structure of $S=-2$ hypernuclei}\\
E.~Hiyama}

%





\titl{Five- and four-body structure of
$S=-2$ hypernuclei}

\name{
E.~Hiyama
}

\adr{
RIKEN Nishina Center, RIKEN, Hirosawa 2-1, Wako, Saitama, 351-0198, Japan}


It is interesting to study the structure of  multi-strangeness
system which consists of many hyperons and nucleons.
In this meaning, the sector of $S=-2$ nuclei, double
$\Lambda$ hypernuclei and $\Xi$ hypernuclei is just the 
entrance to the multi-strangeness world.
However, we have hardly any knowledge of
the $YY$ interaction because there exist no $YY$ 
scattering data.
Then, in order to understand the
$YY$ interaction, it is crucial
to study the structure of double $\Lambda$ hypernuclei
and $\Xi$ hypernuclei.

In KEK-E373 emulsion experiments, there
were observed several events corresponding to
double-$\Lambda$ hypernuclei.
First, observation of NAGARA event  has been reported.
The two $\Lambda$ separation energy, $B_{\Lambda \Lambda}$
was $6.91 \pm 0.16$ MeV [1].
This event was identified uniquely as 
$^6_{\Lambda \Lambda}$He in the ground state.
A second important observation
was the Deamchi-Yanagi event [2,3]
identified as $^{10}_{\Lambda \Lambda}$Be
with $B_{\Lambda \Lambda}=11.09 \pm 0.13$ MeV.
though it was uncertain whether this event was interpreted to be
the ground state or an excited state.
Furthermore, a newly observed
double-$\Lambda$ event has been recently reported, called Hida event [2].
This event has two possible interpretations:
One is $^{11}_{\Lambda \Lambda}$Be with $B_{\Lambda \Lambda}=
20.83 \pm 1.27 $ MeV, and
the other is $^{12}_{\Lambda \Lambda}$Be
with $B_{\Lambda \Lambda}=22.48 \pm 1.21$ MeV.
It is uncertain whether this is an observation of
a ground state or an excited state.

The one of the aim of the present work is
to interpret Demachi-Yanagi event and new  Hida event
within the framework of $\alpha \alpha \Lambda \Lambda$
and $\alpha \alpha n \Lambda \Lambda$ four- and five-body model,
respectively.
The calculated $B_{\Lambda \Lambda}$ is $11.83$ MeV in 
$^{10}_{\Lambda \Lambda}$Be, which is consistent with
the Demachi-Yanagi event. Then, we can
interpret 
this event is an observation of the $2^+$ state
excited state of $^{10}_{\Lambda \Lambda}$Be.
In our five-body calculation, we employ
the interactions of Ref.[4] so that those severe 
constraints are also successfully
met in our two-, three-, and four-body subsystems.
The calculated values of $B_{\Lambda \Lambda}$
is $18.23$ MeV for the
$3/2^-$ ground state, while for the excited
states the $B_{\Lambda \Lambda}$ values
are calculated to be less than $15.5$ MeV.
Therefore, the observed Hida event can be
interpreted to be
the ground state.
This calculated value does not
contradict the 
data within $2 \sigma$.

 For the study of $\Xi N$ interaction, it is important to study
 the structure of $\Xi$ hypernuclei.
 However, so far there is no observed $\Xi$
 hypernucleus.
 Then, it is important to predict theoretically what kinds
 of $\Xi$ hypernuclei will exist as bound states.
 For this purpose, they are planning to
 produce the $^{12}_{\Xi^-}$Be using $^{12}$C target by
 $(K^-,K^+$) reaction. However, even if we observe this system as
 a bound state, we shall get only information
 that $V_{\Xi N}$ itself is attractive.
 Then, next, after this experiment, we want to
 know desirable strength of the spin-, iso-spin independent term.
 For this purpose, we propose to observe the bound states in
 future $(K^-,K^+)$ experiments using
 $^7$Li and $^{10}$B targets.

\vfill  

\noindent{\bf References }
\begin{description}
\setlength\itemsep{-3pt}
\item{[1]} Takahashi {\it et al.}, Phys. Rev. Lett. {\bf 87} (2001) 212502.
\item{[2]} K. Nakazawa {\it et al.},
Nuclear Physics (The proceedings on the
10th International Conference on  on Hypernuclear 
and Strange Particle Physics (Hyp X), Tokai, Sept.14-18, 2009),
in poress and to be submitted to Phys. Rev. C:
K. Nakazawa, private communication (2010).
\item{[3]} A. Ichikawa, PhD Thesis, Kyoto University, 2001.
\item{[4]} E. Hiyama, M. Kamimura,
T. Motoba, T. Yamada, and Y. Yamamoto,
Phys. Rev. {\bf C66}, 024007 (2002).
\end{description}


\setcounter{equation}{0} 
\setcounter{figure}{0}
\clearpage

\addcontentsline{toc}{section}{
{\bf Toward a realistic $\bar{K}N$-$\pi\Sigma$ interaction}\\
T. Hyodo}

%





\titl{Toward a realistic $\bar{K}N$-$\pi\Sigma$ interaction}

\name{
Tetsuo~Hyodo$^{1}$
}

\adr{
$^1$ Department of Physics, Tokyo Institute of Technology, 
Meguro 152-8551, Japan 
}


We discuss the $\bar{K}N$-$\pi\Sigma$ interaction from the viewpoint of chiral SU(3)$\times$SU(3) symmetry~[1]. Emphasizing the importance of the chiral low energy theorem and unitarity of the scattering amplitude, we construct the $S=-1$ meson-baryon scattering amplitude in chiral coupled-channel approach, where the $\Lambda(1405)$ resonance is dynamically generated. Because of the attractive forces in both $\bar{K}N$ and $\pi\Sigma$ channels, the $\Lambda(1405)$ is described by a superposition of two poles. Although the existence of the lower pole affect to the strength of the $\bar{K}N$ interaction, its position is not precisely determined by the present experimental database. \\

Thus, new constraints are highly demanded for the realistic $\bar{K}N$-$\pi\Sigma$ interaction and the understanding of the $\Lambda(1405)$ resonance. Among others, the $\bar{K}N$ threshold data should be taken into account with the highest priority. So far, the threshold branching ratios of the $K^-p$ to various channels have been determined. In addition, the precise $K^-p$ scattering length will be obtained soon by the SIDDHARTA experiment. With the $\chi^2$ fitting to these data, we can construct the meson-baryon scattering amplitude by chiral coupled-channel approach which can be systematically improved by the higher order contributions. The interaction strength around the $\bar{K}N$ threshold will be determined and the uncertainty in the subthreshold extrapolation will be reduced. An attempt in this direction is outlined~[2]. \\

On top of the $\bar{K}N$ threshold data, the information of the $\pi\Sigma$ channel also serves as an important constraint, especially for the interaction at far below the $\bar{K}N$ threshold. This can be illustrated by the result of Ref.~[3] where  in some cases, the depth of the three-body bound state is not directly related to the energy of the $\Lambda(1405)$, and the strength of the $\pi\Sigma$ attraction plays more important role to determine the mass of the three-body bound state. We discuss the effect of the $\pi\Sigma$ scattering length and effective range to the structure of the $\Lambda(1405)$~[4]. Although the $\pi\Sigma$ scattering length cannot be determined by the scattering experiment, it is possible to extract the scattering length through the threshold cusp effect in the weak decay of $\Lambda_c$~[5] The recent progress in lattice QCD technique can also provide the scattering length~[6]. In this way, the $\pi\Sigma$ threshold information may be determined and will further constrain the dynamics of the $\bar{K}N$-$\pi\Sigma$ system. \\

The author thanks Yoichi Ikeda and Wolfram Weise for fruitful discussion. He is supported by  Global Center of Excellence Program by MEXT, Japan through the Nanoscience and Quantum Physics Project of the Tokyo Institute of Technology
and the Grant-in-Aid for Scientific Research from MEXT and JSPS (No.21840026). 

\vfill  

\noindent{\bf References }
\begin{description}
\setlength\itemsep{-3pt}
\item{[1]} 
  T.~Hyodo and W.~Weise,
  Phys.\ Rev.\  C {\bf 77}, 035204 (2008).
\item{[2]} Y.~Ikeda, T.~Hyodo and W.~Weise,
\textit{in preparation}
\item{[3]} Y.~Ikeda and T.~Sato, Phys.\ Rev.\ C {\bf 76}, 035203 (2007).
\item{[4]} Y.~Ikeda, T.~Hyodo, D.~Jido, H.~Kamano, T.~Sato, K.~Yazaki,
\textit{in preparation}
\item{[5]} T.~Hyodo and M. Oka,
\textit{in preparation}
\item{[6]} Y.~Ikeda, \textit{et al.}, HAL QCD collaboration,
\textit{in preparation}
\end{description}


\setcounter{equation}{0} 
\setcounter{figure}{0}
\clearpage

\addcontentsline{toc}{section}{
{\bf Kaonic $^3$He and $^4$He -- analysis of SIDDHARTA data}\\
T. Ishiwatari {\it et al.}}

%


%



\titl{Kaonic $^3$He and $^4$He -- analysis of SIDDHARTA data}

\name{
T. Ishiwatari$^{1}$,
M.~Bazzi$^{2}$,
G.~Beer$^{3}$
L.~Bombelli$^{4}$,
A.M.~Bragadireanu$^{4,5}$
M.~Cargnelli$^{1}$
G.~Corradi$^{2}$,
C.~ Curceanu$^{2}$,
A.~d'Uffizi$^{2}$,
C.~Fiorini$^{4}$,
T.~Frizzi$^{4}$,
F.~Ghio$^{6}$,
B.~Girolami$^{5}$,
C.~Guaraldo$^{2}$,
R.S.~Hayano$^{7}$,
M.~Iliescu$^{2,5}$,
M.~Iwasaki$^{8}$,
P.~Kienle$^{1,9}$,
P.~Levi~Sandri$^{2}$,
A.~Longoni$^{4}$,
J.~Marton$^{1}$,
S.~Okada$^{2}$,
D.~Pietreanu$^{2}$,
T.~Ponta$^{5}$,
A.~Rizzo$^{2}$,
A.~Romero Vidal$^{2}$,
A.~Scordo$^{2}$,
H. Shi$^{7}$,
D.L.~Sirghi$^{2,5}$,
F.~Sirghi$^{2,5}$,
H.~Tatsuno$^{7}$,
A.~Tudorache$^{5}$,
V.~Tudorache$^{5}$,
O.~Vazquez~Doce$^{2}$,
E.~Widmann$^{1}$,
J.~Zmeskal$^{1}$
}

\adr{
$^1$ Stefan-Meyer-Institut f\"{u}r subatomare Physik, Vienna, Austria\\
$^2$ INFN, Laboratori Nazionali di Frascati, Frascati (Roma), Italy \\
$^3$ Dep. of Phys. and Astro., Univ. of Victoria, Victoria B.C., Canada\\
$^4$ Politechno di Milano, Sez. di Elettronica, Milano, Italy\\
$^{5}$ IFIN-HH, Magurele, Bucharest, Romania\\
$^{6}$ INFN Sez. di Roma I and Inst. Superiore di Sanita, Roma, Italy\\
$^{7}$ Univ. of Tokyo, Tokyo, Japan\\
$^{8}$ RIKEN, The Inst. of Phys. and Chem. Research, Saitama, Japan\\
$^{9}$ Tech. Univ. M\"{u}nchen, Physik Dep., Garching, Germany
}

The first observation of the kaonic $^3$He $3d \to 2p$ transition was made using
a gaseous $^3$He target in the SIDDHARTA experiment [1]. Figure 1 shows the energy
spectrum of the kaonic $^3$He X-rays. The peak around 6.2 keV corresponds to 
\begin{figure}[h]
\begin{minipage}[t]{.47\textwidth}
\centering
  \includegraphics[width=.6\textwidth]{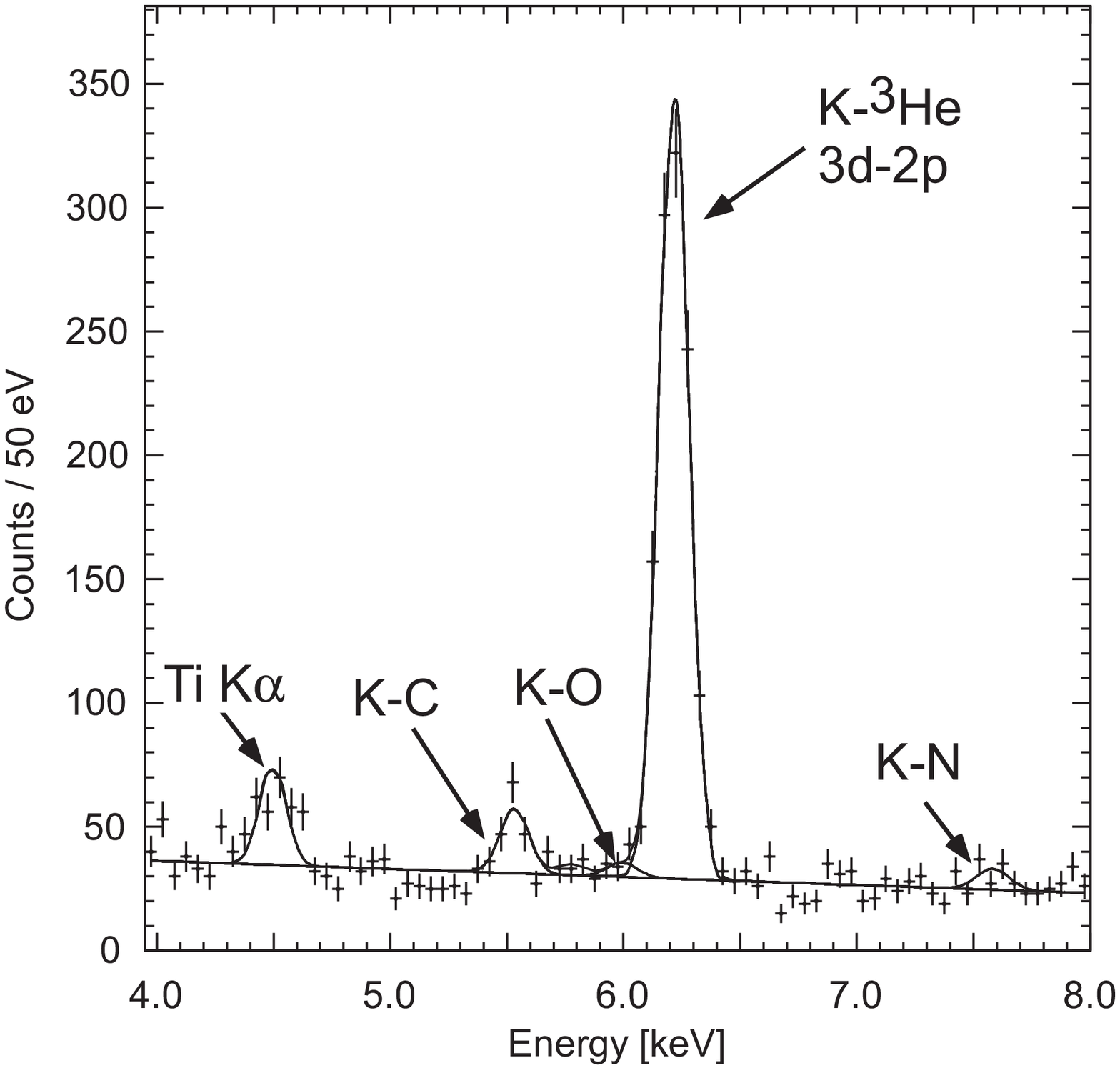}
\caption{X-ray energy spectrum of K-$^3$He.
The $3d \to 2p$ transition energy was determined to
be $6223.0 \pm 2.4 \pm 3.5 $ eV.}
\end{minipage}
\hfill
\begin{minipage}[t]{.47\textwidth}
\centering
  \includegraphics[width=.7\textwidth]{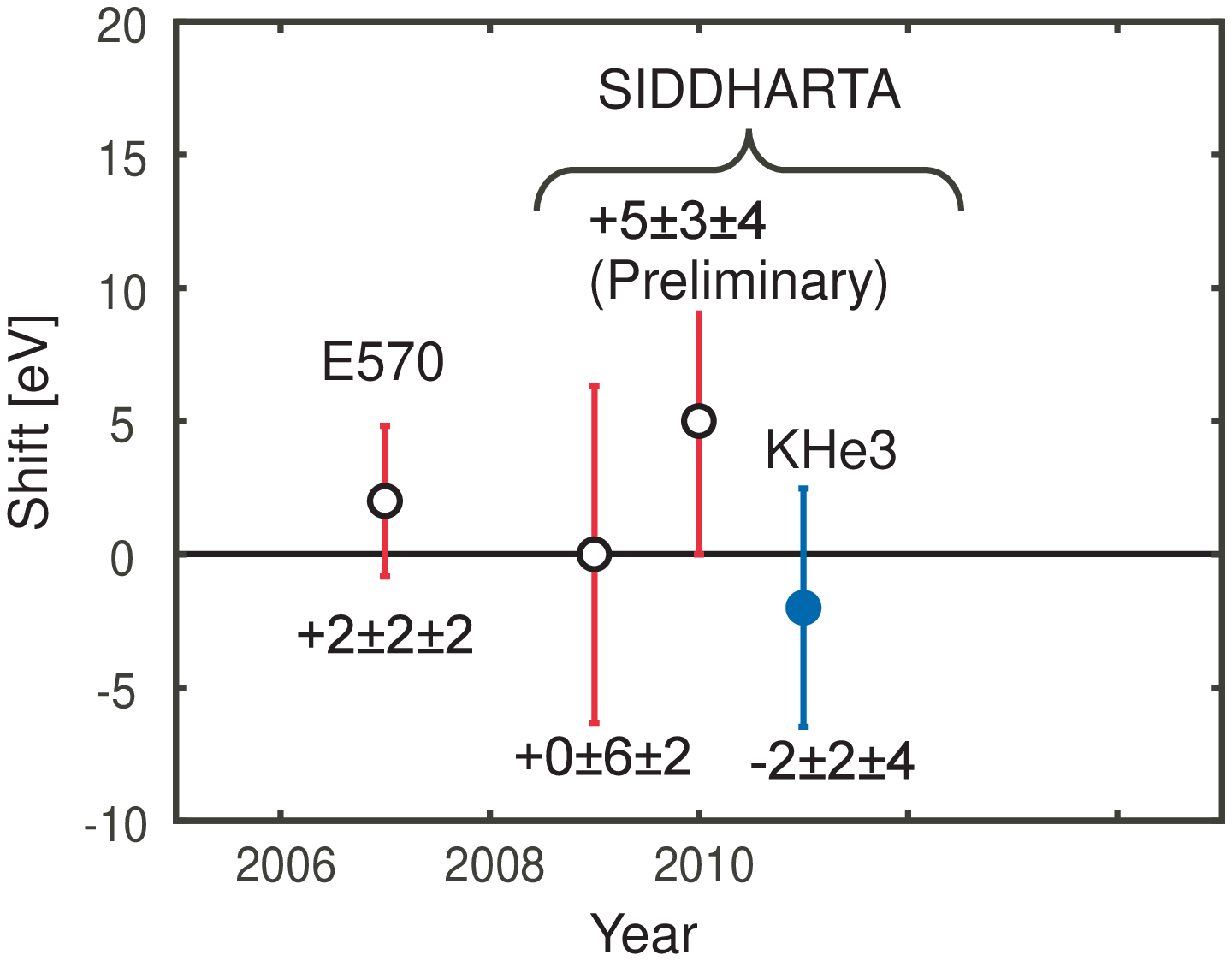}
\caption{Comparison of experimental results. 
Open circle: K-$^4$He $2p$ state, Closed circle: K-$^3$He $2p$ state}
\end{minipage}
\end{figure}
the kaonic $^3$He $3d \to 2p$ transition. 
The strong interaction shift of the kaonic $^3$He $2p$ state was determined to 
be $-2 \pm 2 \mbox{ (stat)} \pm 4 \mbox{ (syst)}$ eV (repulsive). 
A large shift, initially seen in the kaonic $^4$He $2p$ state was not observed.
Experimental results of the shift of the kaonic $^3$He and $^4$He $2p$ states were
summarized in Figure 2.
Together with the recently measured results of the kaonic $^4$He $2p$ state, as well as the
preliminary result of $+5 \pm 3 \mbox{ (stat)} \pm 4 \mbox{ (syst)}$ eV, an abnormal 
shift of the order of 40 eV was not observed
either in kaonic $^3$He or $^4$He [1,2]. However, the size of the experimental errors is still too large
to exclude a possibility of the non-zero shift of the order of 10 eV discussed in [3].

\vfill  

\noindent{\bf References }
\begin{description}
\setlength\itemsep{-3pt}
\item{[1]} SIDDHARTA Collaboration, arXiv:1010.4631v1, submitted to Phys. Lett. B.
\item{[2]} SIDDHARTA Collaboration, M. Bazzi {\it et al.}, Phys. Lett. B {\bf 681} (2009) 310.
\item{[3]} Y. Akaishi, in: Proc. Inter. Conf. on Exotic Atoms (EXA05), Austrian Academy of
Sciences Press, Vienna, 2005, p. 45, http://dx.doi.org/10.1553/exa05s45.

\end{description}


\setcounter{equation}{0} 
\setcounter{figure}{0}
\clearpage

\addcontentsline{toc}{section}{
{\bf Molecule model for kaonic nuclear cluster $\bar{K}NN$, observed by DISTO Collaboration}\\
A.~N.~Ivanov, P.~Kienle, J.~Marton, M.~Pitschmann}

%





\titl{Molecule model for kaonic nuclear cluster $\bar{K}NN$, observed
by DISTO Collaboration}

\name{\underline{\rm A.~N.~Ivanov$^1$}, P.~Kienle$^{2,3}$,
J.~Marton$^2$, M.~Pitschmann$^1$}

\adr{ $^1$ Atominstitut der \"Osterreichischen Universit\"aten,
Technische Universit\"at Wien,\\ Wiedner Hauptstrasse 8-10, A-1040 Wien,
Austria\\ $^2$ Stefan Meyer Institut f\"ur subatomare Physik
\"Osterreichische Akademie der Wissen-\\schaften, Boltzmanngasse 3,
A-1090, Wien, Austria\\ $^3$ Excellence Cluster Universe Technische
Universit\"at M\"unchen, D-85748 Garching,\\ Germany }

The molecule model of kaonic nuclear clusters (KNCs) [1] is extended
to the analysis of the experimental data by the DISTO Collaboration
[2]. Our analysis of the KNCs ${^1_{\bar{K}}}{\rm H}$ and
${^2_{\bar{K}}}{\rm H}$, carried out in [1], is based on the
assumptions that 1) the masses of the KNCs ${^1_{\bar{K}}}{\rm H}$ and
${^2_{\bar{K}}}{\rm H}$ have a structure $M_{{^1_{\bar{K}}}{\rm H}} =
m_N + m_K - B^{WT}_{{^1_{\bar{K}}}{\rm H}}$ and $M_{{^2_{\bar{K}}}{\rm
H}} = 2 m_N + m_K - B^{WT}_{{^2_{\bar{K}}}{\rm H}}$, where the binding
energies $B^{WT}_{{^1_{\bar{K}}}{\rm H}}$ and
$B^{WT}_{{^2_{\bar{K}}}{\rm H}}$ are calculated in the
tree--approximation and defined by the Weinberg--Tomozawa interactions
only, and 2) the stiffnesses of harmonic oscillator potentials,
keeping the pairs $(\bar{K}N)_{I = 0}$ and $N\otimes (\bar{K}N)_{I =
0}$ bound, are equal.  This provides a relation
$\mu_{\Lambda^*}\Omega^2_{\Lambda^*} =
\mu_{\Lambda^*N}\Omega^2_{\Lambda^*N}$ between frequencies
$\Omega_{\Lambda^*}$ and $\Omega_{\Lambda^*N}$, parameterising the
harmonic oscillator wave functions of KNCs and describe correlations
of $\bar{K}N$ and $N(\bar{K}N)$ pairs, having the reduced masses
$\mu_{\Lambda^*}$ and $\mu_{\Lambda^*N}$, respectively. These
assumptions are dropped in the present analysis.  For the description
of the experimental data on the quasi--bound $\bar{K}NN$ state by the
DISTO Collaboration in the molecule model we treat it as the KNC
${^2_{\bar{K}}}{\rm H}$ and assume that the masses of the KNC
${^1_{\bar{K}}}{\rm H}$, which we identify with the $\Lambda(1405)$
resonance (or $\Lambda^*$) with mass $m_{\Lambda^*} = 1405\,{\rm MeV}$
[1], and the KNC ${^2_{\bar{K}}}{\rm H}$ have the structures
$M_{{^1_{\bar{K}}}{\rm H}} = m_N + m_K - B^{WT}_{{^1_{\bar{K}}}{\rm
H}} + \delta M_{{^1_{\bar{K}}}{\rm H}}$ and $M_{{^2_{\bar{K}}}{\rm H}}
= 2 m_N + m_K - B^{WT}_{{^2_{\bar{K}}}{\rm H}} + \delta
M_{{^2_{\bar{K}}}{\rm H}}$. The mass--corrections $\delta
M_{{^1_{\bar{K}}}{\rm H}}$ and $\delta M_{{^2_{\bar{K}}}{\rm H}}$ are
given by the real parts of Feynman self--energy diagrams of KNCs
${^1_{\bar{K}}}{\rm H}$ and ${^2_{\bar{K}}}{\rm H}$, the imaginary
parts of which define the partial widths for the kinematically allowed
decay modes. The mass--corrections are calculated with the account for
all intermediate states of elastic and inelastic $\bar{K}N$ and
$\bar{K}NN$ interactions, described by chiral Lagrangians with
derivative meson--baryon couplings invariant under chiral $SU(3)\times
SU(3)$ symmetry.  For the frequencies $\Omega_{\Lambda^*} = 59.5\,{\rm
MeV}$ and $\Omega_{\Lambda^*N} = 105\,{\rm MeV}$, providing the
optimal ratio $B_{{^2_{\bar{K}}}{\rm H}}/\Gamma_{{^2_{\bar{K}}}{\rm
H}}= 0.83$ in comparison with the experimental one
$B^{(\exp)}_{{^2_{\bar{K}}}{\rm
H}}/\Gamma^{(\exp)}_{{^2_{\bar{K}}}{\rm H}}= 0.87(11)$, we get 1)
binding energies $B_{{^1_{\bar{K}}}{\rm H}} = 29\,{\rm MeV}$ and
$B_{{^2_{\bar{K}}}{\rm H}} = 118\,{\rm MeV}$, 2) widths
$\Gamma_{{^1_{\bar{K}}}{\rm H}} = 36.4\,{\rm MeV}$ and
$\Gamma_{{^2_{\bar{K}}}{\rm H}} = \Gamma_{N\Lambda^0} +
\Gamma_{N\Sigma} = 142\,{\rm MeV}$ and 3) densities
$n_{{^1_{\bar{K}}}{\rm H}}(0) = 0.37\,n_0$ and $n_{{^2_{\bar{K}}}{\rm
H}}(0) = 2.71\,n_0$, where $\Gamma_{N\Lambda^0} = 97\,{\rm MeV}$ and
$\Gamma_{N\Sigma} = 45\,{\rm MeV}$ are the partial widths of the decay
modes ${^2_{\bar{K}}}{\rm H} \to N\Lambda^0$ and $N\Sigma$,
respectively, and $n_0 = 0.17\,{\rm fm^{-3}}$ is the normal nuclear
density. Using the wave functions of KNCs ${^1_{\bar{K}}}{\rm H}$ and
${^2_{\bar{K}}}{\rm H}$ [1], we show that the probability of the
coalescence of the $\Lambda^*N$ pair into ${^2_{\bar{K}}}{\rm H}$ is
$P(\Lambda^*N \to {^2_{\bar{K}}}{\rm H})\sim 0.94$. This allows to
argue that the KNC ${^2_{\bar{K}}}{\rm H}$ with the binding energy
$B_{{^2_{\bar{K}}}{\rm H}} = 118\,{\rm MeV}$ and the width
$\Gamma_{{^2_{\bar{K}}}{\rm H}} = 142\,{\rm MeV}$ can be, in
principle, identified with the compact $\bar{K}NN$ system, measured by
the DISTO Collaboration [2].

\vfill  

\noindent{\bf References }
\begin{description}
\setlength\itemsep{-3pt}
\item{[1]} M. Faber {\it et al.}, arXiv: 0912.2084v2 [nucl-th], J. of
Mod. Phys. E (to be published).
\item{[3]} T. Yamazaki {\it et al.}, (DISTO Collaboration),
Phys. Rev. Letters {\bf 104}, 132502 (2010).
\end{description}


\setcounter{equation}{0} 
\setcounter{figure}{0}
\clearpage

\addcontentsline{toc}{section}{
{\bf Role of $\Lambda$(1405) in the formation of $X(K^-pp)$ from the energy dependence of the 
$pp \to X + K^+$ reaction at 2.50 GeV and  2.85 GeV incident energy}\\
Paul Kienle, In behalf of the DISTO Collaboration}

%





\titl{Role of $\Lambda$(1405) in the Formation of 
$X(K^-pp)$ from the Energy Dependence of the 
$pp \to X + K^+$ Reaction at 2.50 GeV and  2.85 GeV Incident Energy}
\name{
Paul Kienle$^{1}$,
In behalf of the DISTO Collaboration
}

\adr{
$^1$ Excellence Cluster Universe Technische Universit\"{a}t M\"{u}nchen
}

We have analyzed recently data [1] of the DISTO experiment on the exclusive 
$p p \to  p \Lambda K^+$ reaction and found in the $K^+$ missing mass and 
the $p \Lambda$ invariant mass spectra, at high transverse momentum transfer of 
the $p$ and $K^+$, show a broad distinct peak, $X$, of 26-$\sigma$  confidence 
with a mass $M_X = 2267 \pm 3 \mbox{ (stat)} \pm 5 \mbox{ (syst)}$ MeV/c$^2$ and 
a width $\Gamma_X = 118  \pm 8 \mbox{ (stat)} \pm 10 \mbox{ (syst)}$ MeV. 
The large formation probability of $X$, in order of the $\Lambda(1405)$ cross section, 
indicates the formation of a compact $K^-pp$ bound system via a 
$p \Lambda(1405)$ doorway state with a large binding energy of $B_X = 103$ MeV 
which can be a possible gateway toward cold and dense kaonic nuclear matter. 
Here we report on the analysis of DISTO data on the exclusive 
$p p \to p \Lambda K^+$ reaction at 2.50 GeV incident energy to study directly 
the $p \Lambda(1405)$ doorway state hypothesis for the formation of the deeply bound
$M_X(2265)$ state. The population of the $M_X(2265)$ is kinematically allowed at 2.50 GeV 
incident energy and is expected to be produced as much as 50\% of that at 2.85 GeV. 
On the other hand we expect and also verify experimentally the very weak population of 
the $\Lambda(1405)$ state at 2.50 GeV which is less than 10\% of that at 2,85 GeV as 
expected kinematically because of the small energy above the formation threshold.

\begin{figure}[h]
\begin{center}
  \includegraphics[height=.27\textheight]{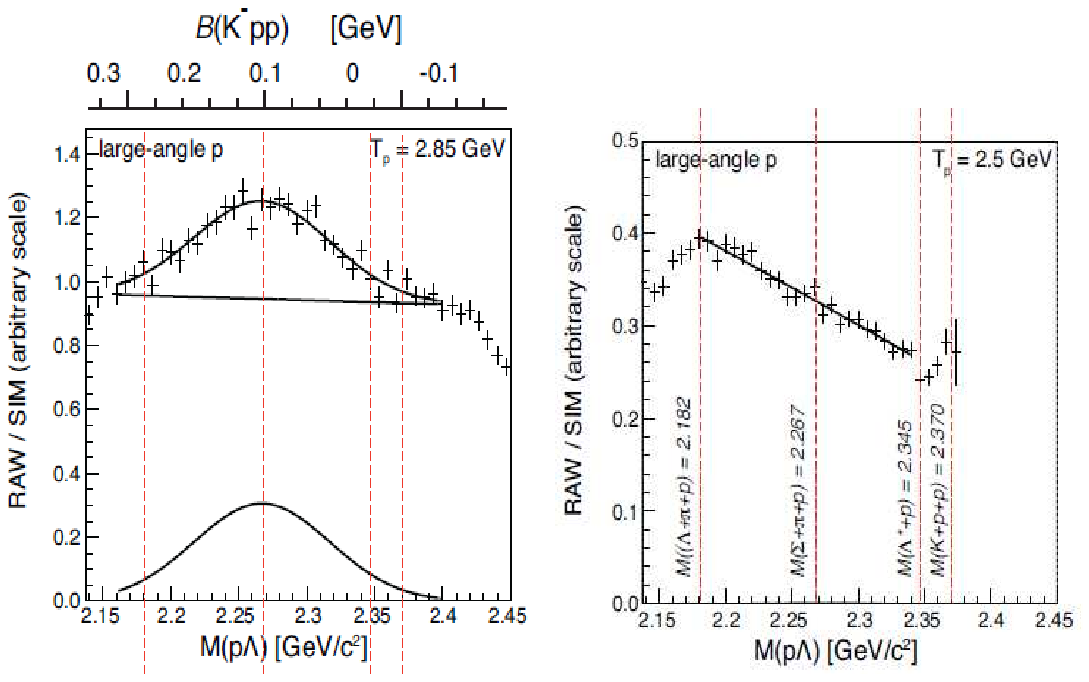}
  \caption{$M(p \Lambda)$ invariant mass spectra at 2.85 GeV (left) and 2.50 GeV (right)}
\end{center}
\end{figure}

A comparison of the $M(p \Lambda)$ invariant mass spectra at 2.85 GeV (left) and 2.50 GeV (right) show 
that the ratio of the cross sections for the formation of $X(2265)$ at 2.50 and 2.85 GeV is  
$X(2.50)/ X(2.85) = 0.011 \pm 0.106$. The absence of $X$ at 2.50 GeV is in line with the very 
weak production of $\Lambda(1405)$ at 2.50 GeV, thus giving direct evidence that $\Lambda(1405)$ is 
the doorway state for the formation of $K^-pp$ with a high $\Lambda(1405)-p$ sticking probability 
at 2.85 GeV energy.

\vfill  

\noindent{\bf References }
\begin{description}
\setlength\itemsep{-3pt}
\item{[1]} T. Yamazaki, M- Maggiora, P. Kienle, K. Suzuki, {\it et al.}, Phys. Rev. Lett, {\bf 104}, 132502 (2010).
\end{description}


\setcounter{equation}{0} 
\setcounter{figure}{0}
\clearpage

\addcontentsline{toc}{section}{
{\bf  $\mathbf {\Lambda(1405)}$: A new hope. Measurements with the pion beam}\\
K.~Lapidus for the HADES collaboration}

%





\titl{ $\mathbf {\Lambda(1405)}$: A New Hope.\\ Measurements with the pion beam}

\name{
K.~Lapidus for the HADES collaboration
}

\adr{
Excellence Cluster ``Universe'', TU M\"unchen, Boltzmannstr. 2, 85748 Garching, Germany

}


The ambiguity in the current theoretical understanding of the $\Lambda(1405)$ resonance calls for new experimental efforts.
It is predicted [1] that the observed properties of the resonance (i.e. position of the pole and the lineshape) depend on both the production mechanism ($\gamma$-, $\pi$-, kaon- or proton-induced reactions) and the decay mode ($\Sigma^{\pm}\pi^{\mp}$, $\Sigma^{0}\pi^{0}$). However, a recent photoproduction measurement contradicts theoretical expectations [2].

Existing bubble-chamber measurements of the $\Lambda(1405)$ production in pion-induced reactions [3-5] have principal flaws. First of all, the neutral decay channel $\Lambda(1405) \rightarrow \Sigma^{0}\pi^{0}$ was never measured. Besides that, the $\Sigma(1385)^{0}$ contribution was not subtracted from the observed spectra. 

Therefore, we propose to measure the $\Lambda(1405)$ in pion-proton reactions ($\pi^{-}p \rightarrow \Lambda(1405)K^{0}$) with the High Acceptance Dielectron Spectrometer at the GSI Helmholtzzentrum.

Since at near-threshold energies the $\Lambda(1405)$ is produced together with a $K^{0}$, the reconstruction of $K^{0}_{S}$, decaying to $\pi^{+}\pi^{-}$ pairs, allows to reconstruct the $\Lambda(1405)$  via missing mass technique. However, exclusive measurements are necessary in order to disentangle the different decay channels, including the neutral decay mode $\Lambda(1405) \rightarrow \Sigma^{0} \pi^{0}$ with subsequent decays $\Sigma^{0} \rightarrow \Lambda \gamma$, $\pi^{0} \rightarrow \gamma\gamma$.

Though in the present configuration the HADES setup is not able to detect photons, the installation of an electromagnetic calorimeter is foreseen: it is planned to use lead-glass blocks from the end-cap calorimeter of the former OPAL experiment. In total, $\sim $900 blocks will be installed in six sectors. Dedicated beam tests showed that individual calorimeter modules offer an energy resolution of $\sigma_{E}/\sqrt{E/GeV} \approx 5\%$, typical for such type of calorimeter.

First studies of the acceptance and achievable mass resolution for the $\Lambda(1405)$ were performed with recently developed simulation and reconstruction software. They show that the HADES setup equipped with the electromagnetic calorimeter will allow, for the first time, for an exclusive reconstruction of the $\Lambda(1405)$ resonance in pion-induced reactions in all of its decay channels.

\vfill  

\noindent{\bf References }
\begin{description}
\setlength\itemsep{-3pt}
\item{[1]} V.~K.~Magas, E.~Oset and A.~Ramos,
  Phys.\ Rev.\ Lett.\  {\bf 95} (2005) 052301.
\item{[2]} K.~Moriya and R.~Schumacher,
  AIP Conf.\ Proc.\  {\bf 1257} (2010) 557.
\item{[3]} G.~Alexander, G.~R.~Kalbfleisch, D.~H.~Miller and G.~A.~Smith,
  Phys.\ Rev.\ Lett.\  {\bf 8} (1962) 447.
\item{[4]} A.~Engler, H.~E.~Fisk, R.~W.~Kraemer, C.~M.~Meltzer and J.~B.~Westgard,
  Phys.\ Rev.\ Lett.\  {\bf 15} (1965) 224.
\item{[5]} D.~W.~Thomas, A.~Engler, H.~E.~Fisk and R.~W.~Kraemer,
  Nucl.\ Phys.\  B {\bf 56} (1973) 15.

\end{description}


\setcounter{equation}{0} 
\setcounter{figure}{0}
\clearpage

\addcontentsline{toc}{section}{
{\bf  Antikaons in nuclei}\\
J. Mare\v{s}}

%





\titl{Antikaons in Nuclei}

\name{J. Mare\v{s}
}

\adr{
Nuclear Physics Institute ASCR, 250 68 \v{R}e\v{z}, Czech Republic 
}

This contribution reviews our recent calculations of the anticipated $K^-$ binding energies $B_{K^-}$
and corresponding conversion widths  $\Gamma_{K^-}$  of the $K^-$ nuclear states. 
Mechanisms that determine the width of the $K^-$ nuclear states are discussed for the relativistic mean field 
model (RMF) with a phenomenological optical model imaginary potential Im$V_{\rm opt}$ [1] and for models based on chiral dynamics 
with coupled channels ('Chiral') [2]. The calculations are performed selfconsistently in order to properly evaluate 
the dynamical effects of $K^-$ on the nuclear core and vice versa. 
In the RMF approach the energy dependence of Im$V_{\rm opt}$ is introduced via a suppression factor, which takes into account  
the reduced phase space available for $K^-$ absorption from deeply bound states [3]. On the contrary, in the 'Chiral' models   
the energy dependence (as well as a weak density dependence) enters directly the scattering amplitude calculated 
within the multichannel framework [2]. The $K^-$ widths are then a result of a delicate interplay between the energy and 
density dependence. Surprisingly enough, even the 'Chiral' models with selfenergies (i.e. with medium modified hadron 
propagators), which yield shallow $K^-$ potentials Re$V_{\rm opt}$ at threshold, predict deep $K^-$ potentials  
in selfconsistent, dynamical calculations of kaonic nuclei. This is illustrated  
 for the Weinberg-Tomozawa (WT) and next-to-leading-order(CS30) models in Table 1 
(see [3] for details). It is to be stressed that the nonmesonic decay mode ${\bar K}NN \rightarrow YN$ was not considered in the table.
       
\begin{table}[h!]
\caption{Potentials $V_{\rm opt}$ in $^{16}$O+$K^-$ for $r=0$ calculated using 'Chiral' models WT and CS30 [2] (in MeV) without and with hadron selfenergies (SE), 
at threshold (first line) and selfconsistently (2nd line).} 
\label{tab1}
\begin{center}
\begin{tabular}{lcccc}
\hline \hline
 & WT &  CS30 &  WT$_{\rm SE}$ &  CS30$_{\rm SE}$ \\ \hline
$V_{\rm opt}(r=0,\;B_{K^-}=0)$ & -62.6-i62.0 & -80.0-i64.3 & -24.7-i50.5 &  -40.2-i61.4 \\ 
$V_{\rm opt}(r=0,\;B_{K^-} \leftarrow SC)$ & -95.0-i28.6 & -114.3 -i16.8 & -97.1-i13.5 & -106.2-i9.9 \\ 
\hline
\end{tabular}
\end{center}
\end{table}
    
As a result, the 'Chiral' models, WT as well as CS30, give relatively deeply bound $K^-$ nuclear states with 
$B_{K^-} \approx 70 - 100$~MeV  in heavier nuclei ($A \ge 12$). The non-mesonic ${\bar K}NN \rightarrow YN$ absorption 
channel contributes significantly to the widths. Our preliminary calculations predict $\Gamma_{K^-} \ge 40$~MeV  
when the non-mesonic absorption mode is included.     

\bigskip
I would like to thank my collaborators E. Friedman, A. Gal, D. Gazda and A. Ciepl\'{y}, with whom 
these calculations were performed. \\
The work was supported by the GA\v{C}R grant 202/09/1441.

\vfill  

\noindent{\bf References }
\begin{description}
\setlength\itemsep{-3pt}
\item{[1]} D. Gazda, E. Friedman, A. Gal, J. Mare\v{s}, Phys. Rev. C  {\bf 76} (2007) 055204.
\item{[2]} A. Ciepl\'y, J. Smejkal, Eur. Phys. J. A {\bf 43} (2010) 191. 
\item{[3]} J. Mare\v{s}, E. Friedman, A. Gal, Nucl. Phys. A {\bf 84} (2006) 770.
\end{description}


\setcounter{equation}{0} 
\setcounter{figure}{0}
\clearpage

\addcontentsline{toc}{section}{
{\bf  Status of the HadronPhysics2 LEANNIS network and the proposed LEANNIS network in HadronPhysics3}\\
J. Mare\v{s}}

%





\titl{Status of the HadronPhysics2 LEANNIS Network and the proposed LEANNIS network in HadronPhysics3}

\name{J. Marton$^1$
}

\adr{
$^1$ Stefan Meyer Institute, Austrian Academy of Sciences, 1090 Vienna, Austria
}


In January 2009 the network LEANNIS (Low-energy Antikaon Nucleon and Nucleus Interaction Studies) - a networking activity within the European integrated activity project HadronPhysics2 [1] went into operation. The remarkable progress in the first 18 months was reported at this meeting.
 Many scientific topics of this ECT$^{*}$ meeting and the previous meeting in 2009 [2] have very close connections to LEANNIS: Precision studies of kaonic atoms like kaonic hydrogen, deuterium and helium isotopes and the question of kaonic nuclear clusters. The network brings together experimentalists and theoreticians working in the field in order to optimize experimental approaches and to interpret the experimental results in the framework of refined theoretical descriptions.\\

Several highlights of LEANNIS were presented. One of the most important successes was the measurement of the Lyman x-ray spectrum of kaonic hydrogen and the first time measurement of the Balmer x-ray spectrum of kaonic $^{3}$He. These experimntal data are important imput for theoretical studies on the strong interaction antikaon-nucleon at threshold. \\

The proposed LEANNIS network in HadronPhysics3 will be concentrated on several hot topics in low-energy antikaon interactions on the nucleon and nuclei to be studied in theory and experiment. It will take advantage of the basis created in the running of the LEANNIS network-project and of its outcome. The research topics include
\begin{itemize}
\item Precision data on the observables of strong interaction in hadronic atoms like kaonic deuterium and heavier kaonic atoms in order to provide a full data set for theoretical studies,

\item Experimental informations from antikaonic nuclear cluster searches to answer the question of their existence, production mechanism and characteristics serving as a crucial input for theory,

\item Experiments and theoretical studies on the nature of sub-threshold strangeness resonances.

\end{itemize}

 LEANNIS will further intensify the close collaboration between theory and experimental groups to explore the fascinating characteristics of the low-energy antikaon interaction, with the aim of an improved understanding of the non-perturbative QCD.\\

\vfill  


\noindent{\bf References }
\begin{description}
\setlength\itemsep{-3pt}
\item{[1]} http:/hadronphysics.infn.it
\item{[2]} C. Curceanu, J. Marton, "Mini-Proceedings ECT* Workshop Hadronic Atoms and Kaonic Nuclei - Solved Puzzles, Open Problems and Future Challenges in Theory and Experiment", Conference Proceedings , ECT* Workshop 2009, arXiv:1003.2328 [nucl-ex]

\end{description}


\setcounter{equation}{0} 
\setcounter{figure}{0}
\clearpage

\addcontentsline{toc}{section}{
{\bf  Production of hypernuclei in relativistic heavy-ion collisions}\\
I. Mishustin}

%





\titl{Production of hypernuclei in relativistic heavy-ion collisions}
\name{
Igor Mishustin$^{1,2}$
}

\adr{
$^1$ FIAS, Goethe University, 60438 Frankfurt am Main, Germany

$^2$ Russian Research Center Kurchatov Institute, 123182 Moscow, Russia
}

 We have developed a hybrid approach for description of hyper-fragment
  production in the peripheral relativistic heavy-ion collisions [1]: The
  first dynamical stage of the reaction is described by the Quark Gluon
  String Model (QGSM) [2], or by Ultra-relativistic Quantum Molecular
  Dynamics (UrQMD) model [3], which have demonstrated previously a very good
  description of experimental data on strangeness production. We have applied
  a new criterion for absorption of the produced hyperons (Y), mostly
  Lambdas: a hyperon is captured when its kinetic energy is less than its
  potential generated by surrounding spectator nucleons. It is more
  physically justified than a phenomenological coalescence criterion used
  previously [4]. The characteristics of the hyperonic spectator matter and
  probability of producing multi-strange system were also calculated. For
  reliable predictions we need to accumulate about $10^6$ events. The
  calculations are performed at the Center for Scientific Computing, Goethe
  University. We have found that on the level of a few per mile even 3
  Lambdas can be captured simultoneousely in a single spectator.

  At next step the properly modified Statistical Multifragmentation Model
  (SMM) [5] is applied for description of disintegration of strange spectator
  matter into hyper-fragments. It was shown [6] that by comparing theoretical
  predictions with experimental yields of hyper-fragments one can extract new
  information on mass formulae of hyper-nuclei, and estimate the hyperon
  potentials in nuclear matter. We are going to investigate properties of the
  excited hyper-matter, including EOS, caloric curve, and other
  thermodynamical characteristics. A final goal will be a comparison of our
  theoretical predictions with experimental data from GSI and FAIR. This will
  allow us, complementary to nuclear structure studies, to extract
  information about YN and YY interactions, as well as to find out how the
  strangeness admixture will influence properties of nuclear matter in a
  broad range of densities, temperatures, and isospin asymmetries.

\vfill  

\noindent{\bf References }
\begin{description}
\setlength\itemsep{-3pt}
\item{[1]} K.K. Gudima, A.S. Botvina, J. Steinheimer, M Bleicher, I.N. Mishustin,  to be submitted.

\item{[2]} V.D. Toneev and K.K. Gudima, Nucl. Phys. A400 (1983) 173c ; N.S. Amelin, K.K. Gudima, 
and V.D. Toneev, Sov. J. Nucl. Phys. 52 (1990) 1722.

\item{[3]} S.A. Bass et al., Prog. Part. Nucl. Phys. 41 (1998) 255.

\item{[4]} Th. Gaitanos, H. Lenske, U. Mosel, Phys. Lett. B663 (2008) 197; B675  (2009) 297.

\item{[5]} J.P. Bondorf, A.S. Botvina, I.N. Mishustin, et al., Phys. Rep. 257  (1995) 133.

\item{[6]} A.S. Botvina and J. Pochodzalla, Phys. Rev. C76 (2007) 024909.

\end{description}


\setcounter{equation}{0} 
\setcounter{figure}{0}
\clearpage

\addcontentsline{toc}{section}{
{\bf  Status of FOPI@GSI p-p-experiment}\\
R. M\"{u}nzer for the FOPI Collaboration}

%





\titl{Status of FOPI@GSI p-p-experiment}

\name{
R. M\"unzer$^{1}$ for the FOPI Collaboration
}

\adr{
$^1$ Excellence Cluster Universe, Technical University, Munich, Germany
}

In the last years extensive discussions have been carried out about the possible existence of kaonic nuclear clusters, like the $ppK^{-}$ with a predicted mass of M=2322 MeV/$c^{2}$, a binding energy between 20-100 MeV and a width between 40-80  MeV [1,2]. \\
An Experiment was carried out at the FOPI Spectrometer at GSI to investigate the possible production of the $ppK ^{-}$ in a proton-proton reaction at $E_{Kin}=3.1$ GeV, by reconstructing of the reaction (1).
\begin{equation}
p +  p \rightarrow K^{+} + ppK^{-} \rightarrow K^{+} + p + \Lambda \rightarrow K^{+} + p + p + \pi^{-}
\end{equation}
Essential for the reconstruction of this reaction is an efficient identification of $K^{+}$ and the reconstruction of $\Lambda$, especially in the small polar angle region. Figure 1 shows momentum times charge versus the particle velocity measured with the resistive plate chambers of FOPI. The preliminary calibration of the velocity measurement allows a separation of $K^{+}$ from p and $\pi^{+}$ up to momentum of  0.5 $GeV ^{-1}$. With a precise calibration a separation up to 1 $GeVc^{-1}$ should be possible.\\
The upper picture in figure 2 show the invariant mass of p and $\pi^{-}$ emitted to small polar angles (black) together with a polynomial background (red). The lower picture shows the puire signal obtained after background subtraction, in which a clear signal of $\Lambda$ with a width of $\sigma = 12,5 MeVc^{-2}$ can be seen. At the moment refitting methods are under investigation to improve the resolution of the $\Lambda$.
\begin{figure}[h]
\begin{center}
\begin{minipage}[t]{0.46\textwidth}
\center
\includegraphics[width=7.2 cm]{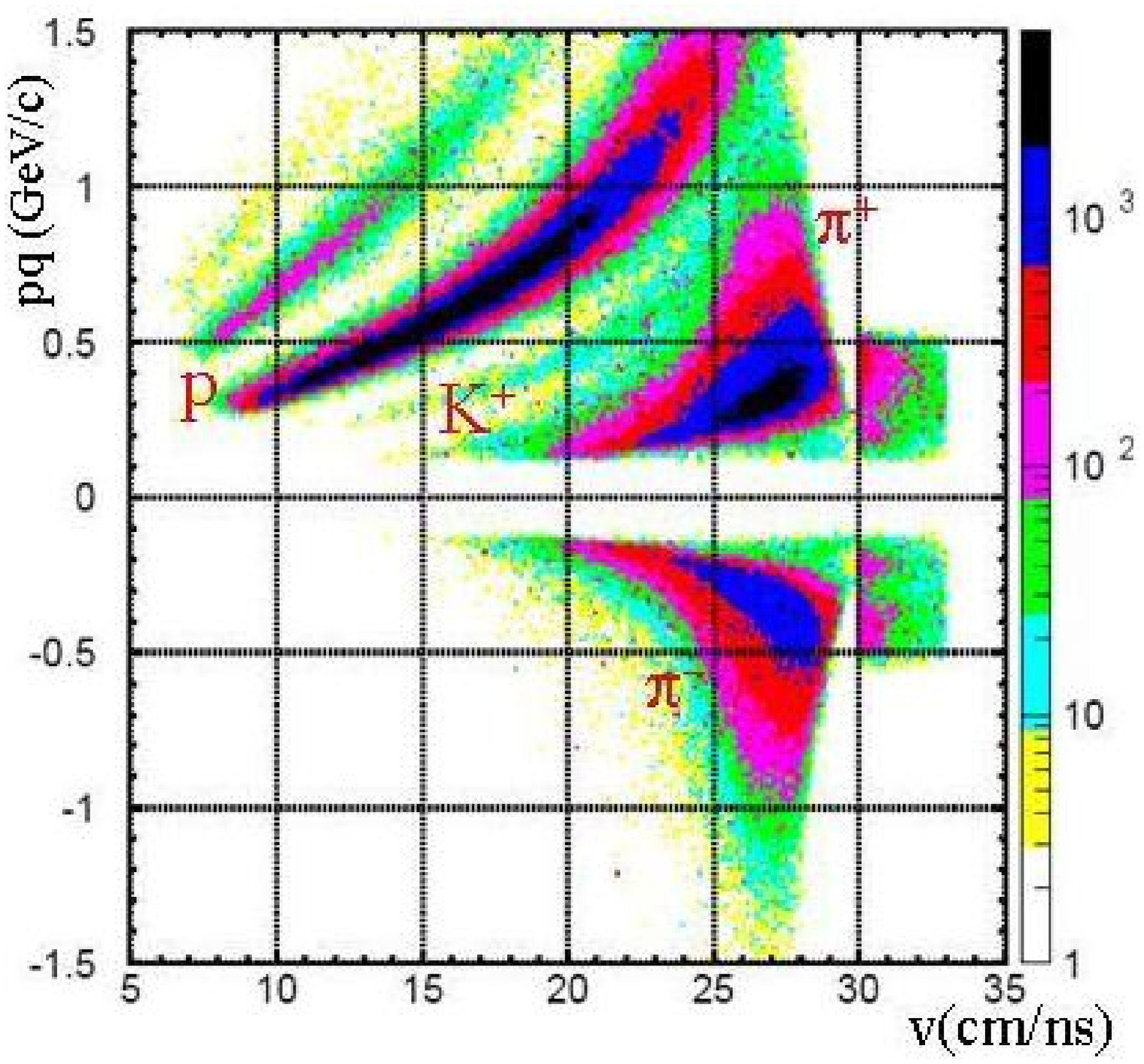}
\caption{\label{figMMpK2} Momentum times charge versus velocity.}
\end{minipage}
\hspace{0.2 cm}
\begin{minipage}[t]{0.46\textwidth}
\center
\includegraphics[width=7. cm]{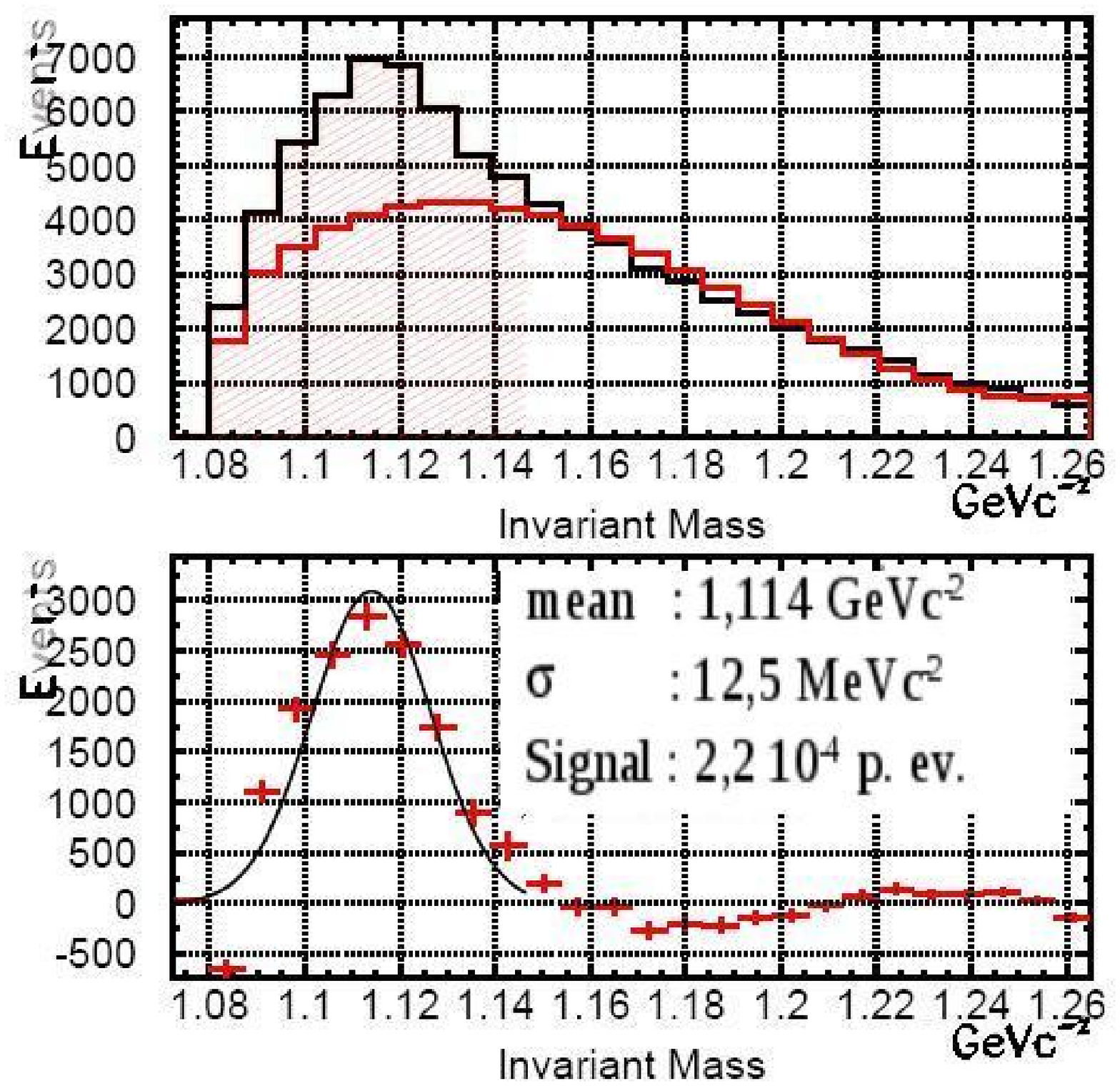}
\caption{\label{lambda} Invariant mass of p and $\pi^{-}$ reconstructed in small polar angles.}
\end{minipage}
\end{center}
\end{figure}
\vfill  

\noindent{\bf References }
\begin{description}
\setlength\itemsep{-3pt}
\item{[1]} T. Yamazaki and Y. Akaishi, Phys. Rev. C {\bf 65} (2002) 70
\item{[2]} T. Dote, T. Hyodo, W.Weise, Phys. Rev. C {\bf 79} (2009) 014003
\end{description}


\setcounter{equation}{0} 
\setcounter{figure}{0}
\clearpage

\addcontentsline{toc}{section}{
{\bf Kaonic hydrogen result from SIDDHARTA}\\
S.~Okada, on behalf of the SIDDHARTA collaboration}

%





\titl{Kaonic hydrogen result from SIDDHARTA}

\name{
S.~Okada$^1$
on behalf of the SIDDHARTA collaboration \\
}

\adr{
$^1$ INFN, Laboratori Nazionali di Frascati, E. Fermi 40, 00044, Frascati (Rome), Italy \\
}


The SIDDHARTA collaboration have performed a measurement of $K$-series
x-rays of kaonic hydrogen atom to determine the strong-interaction
energy-level shift and width of the $1s$ atomic state with significant
improvements over the previous experiments.
The measurement offers a unique possibility to precisely determine
the $\overline{K}$-nucleon scattering lengths which are directly
connected with the physics of the $\overline{K}$N interaction,
and thus were eagerly awaited.
Many studies of the $\overline{K}$-nucleus interaction,
both experimental and
theoretical, (e.g.,deeply-bound K-nucleus state) have been reported
over the past several years. The result will also set tight
constraints on the theories.

\begin{wrapfigure}[20]{r}{8.2cm}
\begin{footnotesize}
\begin{flushleft}
\vspace*{-0.8cm}
\epsfig{file=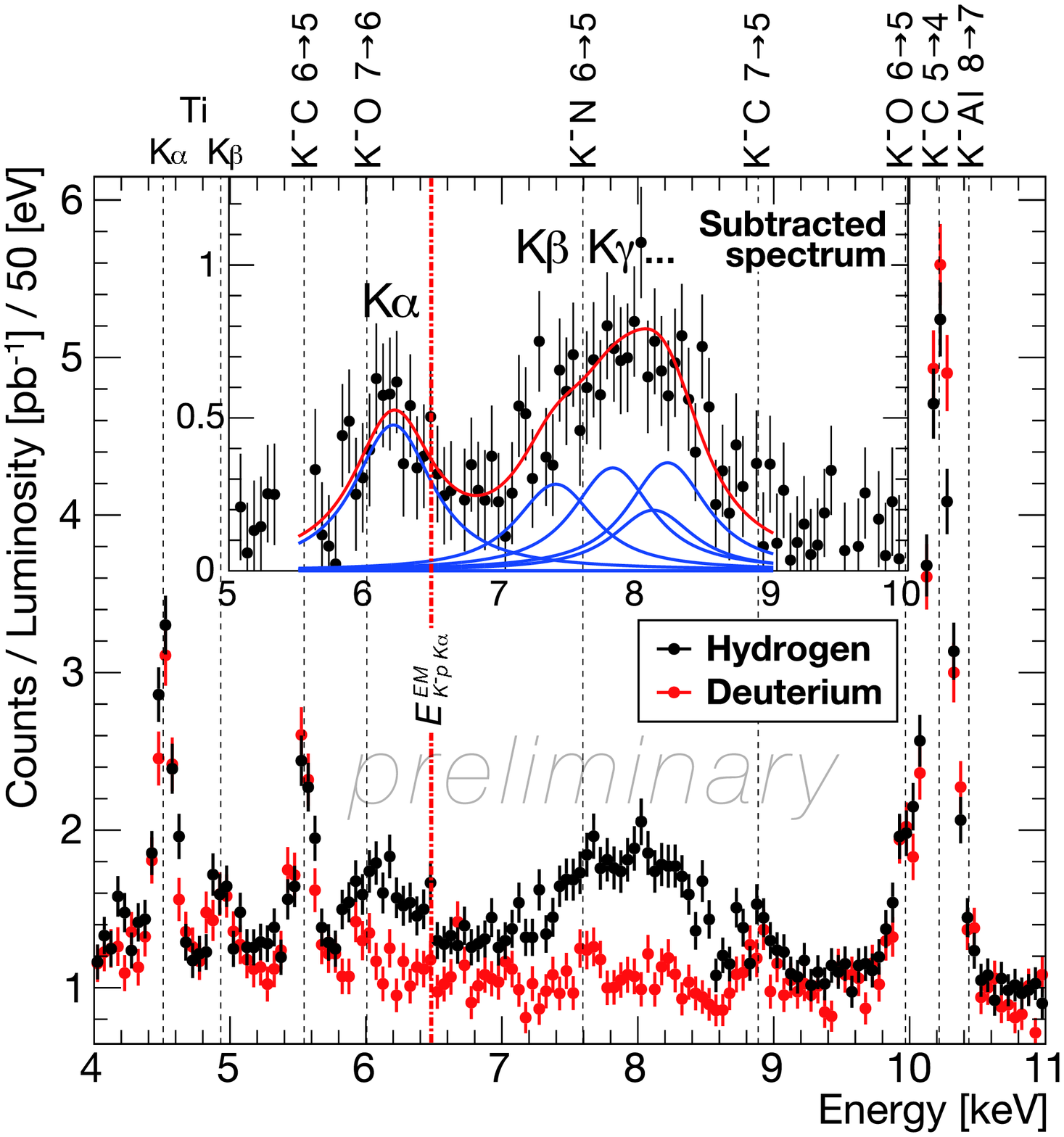,width=8.2cm}
\vspace*{-0.6cm}
\caption{Measured x-ray spectra with hydrogen and deuterium targets.}
\label{fig}
\end{flushleft}
\end{footnotesize}
\end{wrapfigure}

\vspace*{0.1cm}
The experiment was performed at
the DA$\rm{\Phi}$NE positron-electron collider.
A detailed description of our experimental setup
is given in a separate paper [1].

Figure \ref{fig} shows the measured x-ray spectrum
taken with hydrogen target
normalized by the integral luminosity,
overlaying that taken with deuterium target with exactly the same conditions.
$K$-series x-ray transitions of kaonic hydrogen
were clearly observed
while those for kaonic deuterium were hardly seen.
This appears to be consistent with the theoretical expectation of lower
x-ray yield and
greater width for the case of deuterium.
Many other kaonic-atom x-rays were detected in both spectra
as indicated with dotted lines in the figure.
Those lines are attributable to
the target-cell wall made of Kapton
polyimide film (C$_{22}$H$_{10}$O$_5$N$_2$)
and its support frames made of aluminium.

The inset of Fig.\ \ref{fig} shows the kaonic hydrogen x-ray
spectrum after background subtraction
assuming the x-ray spectrum with deuterium target to be backgorund,
and a preliminary fit result
with each fit components.

A dot-dashed line in the figure
indicates the $K_{\alpha}$ x-ray energy
calculated using only the electromagnetic interaction (EM)
for kaonic hydrogen.
Comparing the kaonic-hydrogen $K_\alpha$ peak
and the EM value, there is no room for doubt
about a repulsive-type shift
of the kaonic-hydrogen $1s$-energy level.

The preliminary analysis indicates that 
the statistical accuracies
of the shift and width of kaonic hydrogen
are significantly higher than in previous measurements.
The analysis is now being finalized towards publication.

\vfill  

\noindent{\bf References }
\begin{description}
\setlength\itemsep{-3pt}
\item{[1]} D. Sirghi {\it et al.}, contribution to these Proceedings,
and refrences therein.
\end{description}


\setcounter{equation}{0} 
\setcounter{figure}{0}
\clearpage

\addcontentsline{toc}{section}{
{\bf Studies of $\overline K$ absorption on few nucleon systems and the
search for bound kaonic nuclear states with FINUDA}\\
S. Piano}

%





\titl{Studies of $\overline K$ absorption on few nucleon systems and the
search for bound kaonic nuclear states with FINUDA}

\name{
S. Piano$^{1}$
}

\adr{
$^1$ I.N.F.N. Trieste, via A. Valerio, 2, 34127 Trieste, Italy \\
}



When a $K^-$ is absorbed by few nucleons, a process studied scarcely so far [1],
non-mesonic final states are produced. However, such final states can also be attained in
the decay of possible intermediate aggregates formed by the antikaon and a few nucleons and
bound by the $\overline K N$ attractive force, known as
Bound Kaonic Nuclear States.
According to some Authors [2] the binding energy of such systems is so large
as to prevent their mesonic decay into a $\Sigma\pi$ pair: so they can be
observed as narrow states decaying only in a hyperon and one or more nucleons.
This vision, however, is not shared by a large part of the theoretical community.
In general, much larger widths are expected
which prevent the experimental observation [3].

Several experiments have been performed recently to shed light on this appealing
subject. One of them is FINUDA, that has been able to provide precise measurements
of particle momenta together with $\Lambda$ hyperons identification through
secondary vertices reconstruction.
Back-to-back angular correlations between $\Lambda$'s and protons [4],
deuterons [5] and tritons [6] have been observed, hinting at the existence of some
unexpected dynamics over the standard kaonic absorption mechanism.

The first observation by FINUDA [4] of a bump in the $(\Lambda p)$ invariant mass spectrum
below the Quasi-Free $K^-pp$ absorption reaction threshold 
was based on a first set of data collected on several light targets ($^6$Li, $^7$Li, $^{12}$C).
Exploiting a larger data set the analysis can be performed selecting one target type at a time.
$^6$Li is the lightest target nucleus available in FINUDA.
The contribution of Final State Interactions could spoil significantly the experimental 
observations [7] but it is small in $^6$Li.
The $(\Lambda p)$ invariant mass spectrum from the 
$^6\mathrm {Li}(K^-_{stop}, \Lambda p)X$ reaction
basically confirms the features reported in Ref. [4]. 
The study of the missing mass spectrum allows to single out some Quasi-Free two-nucleon absorption 
reaction channels.
Still after the identification of the contributions from the
 $K^-\ ^6\mathrm {Li}\rightarrow \Lambda(\Sigma^0) p ^4{\mathrm H}$ and
$K^-\ ^6\mathrm {Li}\rightarrow \Lambda\pi^{0,\pm,\mp}\ ^4{\mathrm H}$
reactions, a part of the $(\Lambda p)$ invariant mass spectrum remains unexplained.
Not even the $\Sigma N$ conversion mechanisms can be invoked, as they are unable 
to reproduce the observed angular distributions.

A similar analysis is currently underway on the 
$^6\mathrm{Li}(K^-_{stop},\Lambda d)X$ reaction. The enhancement
observed in Ref. [5] corresponds to a more composite structure which, 
again, cannot be explained by the simple three-nucleon
absorption reactions  $K^-\ ^6\mathrm{Li}\rightarrow \Lambda(\Sigma^0) d\ ^3{\mathrm H}$.

\vfill  

\noindent{\bf References }
\begin{description}
\setlength\itemsep{-3pt}
\item{[1]} P.A. Katz {\it et al., Phys. Rev.} {\bf D1} (1970), 1267;
C. Vander Velde-Wilcquet {\it et al., Nuovo Cim.} {\bf A39} (1977), 538
\item{[2]} Y. Akaishi, T. Yamazaki, {\it Phys. Rev.} {\bf C65} (2002), 044005;
T. Yamazaki, Y. Akaishi,  {\it Nucl. Phys.} {\bf B535} (2002), 70
\item{[3]} E. Oset {\it et al, Nucl. Phys.} {\bf A671} (2002), 481;
J. Mare$\check{\mathrm s}$ {\it et al, Nucl. Phys.} {\bf A770} (2006), 84;
N. Shevchenko {\it et al, Phys. Rev. Lett.} {\bf 98} (2007), 082301;
W. Weise, H. H\"artle, {\it Nucl. Phys.} {\bf A804} (2008), 173;
T. Hyodo, W. Weise, {\it Phys. Rev.} {\bf C77} (2008), 035204
\item{[4]} FINUDA Collaboration, M. Agnello {\it et al., Phys. Rev. Lett.} {\bf 94} (2005), 212303
\item{[5]} FINUDA Collaboration, M. Agnello {\it et al., Phys. Lett.} {\bf B654} (2007), 80
\item{[6]} FINUDA Collaboration, M. Agnello {\it et al., Phys. Lett.} {\bf B669} (2008), 229
\item{[7]} V.K. Magas  {\it \it et al., Phys. Rev.} {\bf C74} (2006), 025206

\end{description}

\vfill

\setcounter{equation}{0} 
\setcounter{figure}{0}
\clearpage

\addcontentsline{toc}{section}{
{\bf A search for double anti-kaon production in antiproton-$^{3}$He annihilation at J-PARC}\\
F.~Sakuma}

%





\titl{A search for double anti-kaon production in antiproton-$^{3}$He
annihilation at J-PARC}

\name{
Fuminori~Sakuma$^{1}$
}

\adr{
$^1$ RIKEN Nishina Center, RIKEN, Wako, 351-0198, Japan
}

 Possible existence of anti-kaonic nuclear clusters has been
investigated extensively both with theoretical and experimental
approaches in recent years.
 In view of the strongly attractive $K^-N$ interaction,
existence of nuclear clusters with more than one $K^-$ is predicted
also, such as $K^-K^-NN$ double anti-kaonic nuclear systems~[1].
 Double anti-kaonic nuclear clusters are predicted to have binding
energies up to 300 MeV and extreme high density exceeding $\sim$ 5 - 6
times than that of the average one $\rho$(0)= 0.17 fm$^3$, thus
producing conditions in the phase diagram of hadronic matter for which
phase transitions to Kaon-condensation / color superconductivity or
pre-cursor effects for these may be reached at low temperature.
 In order to explore such ``strong attraction'' mediated by double
anti-kaons, experimental searches for double anti-kaonic nuclear systems
in $\bar{p}+A$ annihilation have been proposed~[2].

 The elementary anti-proton annihilation reaction, which produces two
pairs of ($K^+ K^-$), is considered as ${\overline{p}} + p \rightarrow
K^+ + K^+ + K^- + K^-  - 98 MeV$ with a negative $Q$-value of 98 MeV,
therefore it is forbidden for stopped antiprotons.
 However, if multi kaonic nuclear cluster exists with deep bound energy
suggested by Ref.~[1], following $\bar p $ annihilation reactions will
be possible on He targets~[2], such as:
\begin{eqnarray}
\label{2.d}
	{\overline{p}} + {^3{\rm He}}
	&\rightarrow&
	K^+ + K^0 + ppK^-K^-  + B_{KK}^{pp} - 109MeV .
\end{eqnarray}
 This double kaonic nuclear cluster process occurs if the binding energy
of the two $K^-$ in a $ppK^-K^-$ cluster $B_{KK}$ exceeds 109 MeV. 
 For the reaction in the final state (\ref{2.d}), we can measure the
missing mass from the $K^+K^0$ energies and also invariant mass of the
decay products, e.g., $ppK^-K^-$ decays into $\Lambda$$\Lambda$.
 For the decay branching-ratio to the $\Lambda \Lambda$ final state, we
needs detailed theoretical evaluation, although this coherent kaon
absorption strength would not be small because of the favored
isospin-zero channel.

 We propose to perform the experiment at the existing K1.8BR beamline at
J-PARC with the E15 spectrometer which consists of a high-precision beam
line spectrometer, a Cylindrical Detector System that surrounds a
target, and a forward neutron TOF counter. 
 The expected yield of stopped $\bar{p}$ is 250 per pulse (3.5s)
with beam momentum of 0.7~GeV/$c$, beam intensity of 6.5$\times$10$^3$
at 50~kW beam power, and a stopping rate of 3.8\%.
 With some assumptions of the double-strangeness production yield and
branching ratio, the proposed experiment has sensitivity to the
$K^-K^-pp$ production yield of 4$\times$10$^{-5}$ per stopped $\bar{p}$
with 3$\sigma$ significance in exclusive measurement of
$K^-K^-\Lambda\Lambda$ final state, in 6 weeks of data taking.
 We can also reach sensitivity to the double-strangeness production
in $\bar{p}+A$ annihilation at rest of 2$\times$10$^{-6}$ per
stopped $\bar{p}$ with 3$\sigma$ significance in inclusive
$\Lambda\Lambda$ measurement, which has never been observed in high
statistics measurements.

\vfill  

\noindent{\bf References }
\begin{description}
\setlength\itemsep{-3pt}
\item{[1]} Y.~Akaishi and T.~Yamazaki, Phys. Rev., {\bf C65} (2002)
		     044005.; Y.~Akaishi, A.~Dote and T.~Yamazaki,
		     Phys. Lett., {\bf B613} (2005) 140.
\item{[2]} W.~Weise, arXiv: nucl-th0507058 (2005).; P.~Kienle,
		     J. Mod. Phys., {\bf A22} (2007) 365.; P. Kienle,
		     J. Mod. Phys., {\bf E16} (2007) 905.; F.~Sakuma
		     {\it et al.}, J-PARC LOI,
		     http://j-parc.jp/NuclPart/pac\_0907/pdf/LOI\_Sakuma.pdf.
\end{description}


\setcounter{equation}{0} 
\setcounter{figure}{0}
\clearpage

\addcontentsline{toc}{section}{
{\bf Coupled-channel Faddeev calculations \\of  $\bar{K}NN - \pi \Sigma N$ system}\\
N.V.~Shevchenko}

%





\titl{Coupled-channel Faddeev calculations \\of  $\bar{K}NN - \pi \Sigma N$ system}

\name{
N.V.~Shevchenko$^{1}$
}

\adr{
$^1$ Nuclear Physics Institute, 25068 \v{R}e\v{z}, Czech Republic
}

Knowledge of $\bar{K}N$ interaction is necessary for
investigation of antikaonic nuclear clusters, attracted large interest
recently. However, the interaction is not known very well. In particular,
there are debates about nature of $\Lambda(1405)$ resonance,
which couples $\bar{K}N$ and $\pi \Sigma$ channels.
Different theoretical groups argue for one- or two-pole structure
of the resonance. We investigated dependence of the $K^- d$ scattering length
on the different models of $\bar{K}N$ interaction, describing $\Lambda(1405)$
resonance in terms of one or two poles.

The $\bar{K}NN - \pi \Sigma N$ system is investigated by coupled-channel
Faddeev equations in AGS form. The inhomogeneous system of equations was solved
in momentum representation in isospin basis.
The two-body $\bar{K}N - \pi \Sigma$ interaction models [1], reproducing all
existing experimental data on $K^- p$ scattering and $K^- p$ atom level shift, were
used with new sets of parameters.
One- and two-pole versions describe all two-body experimental data equally well,
the only difference is in the kaonic hydrogen $1s$ level widths.

We used different versions of the $NN$ potential:
a two-term $NN$ potential [2] reproducing Argonne $NN$ $v18$ phase shifts
(TSA) and therefore having repulsion at short distances and
PEST $NN$ potential from Ref.[3]. Both provide correct
deuteron binding energy, $NN$ scattering length, and effective radius.
A new set of $\Sigma N (-\Lambda N)$ potential parameters
reproducing existing experimental data on $\Sigma N$ and $\Lambda N$
scattering was used. The $I=3/2$ part of the interaction is a one-channel $\Sigma N$,
while $I=1/2$ $\Sigma N$ is connected with $\Lambda N$ channel. Due to this
a coupled-channel $I=1/2$  $\Sigma N - \Lambda N$ potential
was constructed first, then a corresponding one-channel optical
potential for $\Sigma N$ was derived.

The results of $K^- d$ scattering length calculations show, that one- and two-pole
versions of $\bar{K}N - \pi \Sigma$ interaction provide clearly separated sets of
$a_{K^- d}$ values. The scattering length can be used for a calculation of
level shift and width of kaonic deuterium and thereby compared with
experimental results. The difference between $a_{K^- d}$
obtained with one- and two-pole versions of the $\bar{K}N - \pi \Sigma$
potential means, that an accurate measurement of the kaonic deuterium
characteristics, say, in SIDDHARTA experiment [4], could help in
resolving one-/two-pole puzzle of the $\Lambda(1405)$ resonance.

We also repeated our Faddeev calculations of $\bar{K}NN-\pi \Sigma N$ system~[5]
searching for three-body poles in $K^- pp$ with same one- and two-pole
$\bar{K}N - \pi \Sigma$ potentials and other two-body models of interactions
($NN$ and $\Sigma N$) with corresponding quantum numbers. We found, that 
three-body resonance parameters obtained with
one- and two-pole versions of the $\bar{K}N -\pi \Sigma$ potential are different:
two-pole potentials provide slightly more deep and much more
narrow quasi-bound states in the $K^- pp$ system.

More detailed description of the work is in progress.


\vspace{1mm}
\noindent
{\it Acknowledgments.} The work was supported by the Czech GA AVCR grant KJB100480801.

\vfill  

\noindent{\bf References }
\begin{description}
\setlength\itemsep{-3pt}
\item{[1]} J. R\'evai, N.V. Shevchenko, Phys. Rev. {\bf C 79} (2009) 035202.
\item{[2]} P. Doleschall, {\it private communication}.
\item{[3]} H. Zankel, W. Plessas, J. Haidenbauer, Phys. Rev. C 28 (1983) 538.
\item{[4]} C. Curceanu {\it et al}., Eur. Phys. J. A 31 (2007) 537.
\item{[5]} N.V. Shevchenko, A. Gal, J. Mare\v{s}, J. R\'evai, Phys. Rev. {\bf C 76} (2007) 044004.
\end{description}


\setcounter{equation}{0} 
\setcounter{figure}{0}
\clearpage

\addcontentsline{toc}{section}{
{\bf The charged decay channels of the $\pmb{\Lambda(1405)}$ in p+p reactions}\\
J.~Siebenson for the HADES collaboration}

%





\titl{The charged decay channels of the $\pmb{\Lambda(1405)}$ in p+p reactions}

\name{
J.~Siebenson$^{1}$ for the HADES collaboration
}

\adr{
$^1$ Excellence Cluster 'Universe',TU M\"unchen, Boltzmannstr. 2, 85748 Garching, Germany
}


With the predictions of deeply bound kaonic clusters [1], the $\Lambda(1405)$ attracted a lot of interest in the last years. As this particle lies only about 30 MeV/c$^2$ below the kaon nucleon threshold, it is assumed to be an interference of a $\bar{K}N$- bound state and a $\Sigma\pi$- resonance [2]. To disentangle the relative contributions of these two states and to shed light on the still unexplored kaon-nucleon potential, a precise measurement of the $\Lambda(1405)$ line shape is required. We investigated this hyperon excitation by analyzing p+p data at 3.5 GeV kinetic beam energy measured with HADES at GSI ($p+p\rightarrow\Lambda(1405)+p+K^+$). A detailed description of the spectrometer can be found in [3].
The $\Lambda(1405)$ was investigated in the charged $\Sigma^{\pm}\pi^{\mp}$ decay channels, $\Lambda(1405)\rightarrow\Sigma^{\pm}\pi^{\mp}\rightarrow(\pi^{\pm}+n)+\pi^{\mp}$, where also a small contamination of $\Sigma(1385)^0$ is expected. Due to the large acceptance of HADES and its excellent capability for hadron identification, the four charged particles could be reconstructed for a semi-exclusive analysis, whereas the neutron was identified via the missing mass technique. The so obtained data sample was used to reconstruct the two $\Sigma$ hyperons via the missing mass of the proton, the $K^+$ and one of the two pions. After the extraction of the corresponding $\Sigma$ signals, the $\Lambda(1405)$ was investigated via the missing mass of the proton and the $K^+$ separately for both decay channels. 
\begin{figure}[h]
\centering
	\includegraphics[width=0.75\textwidth]{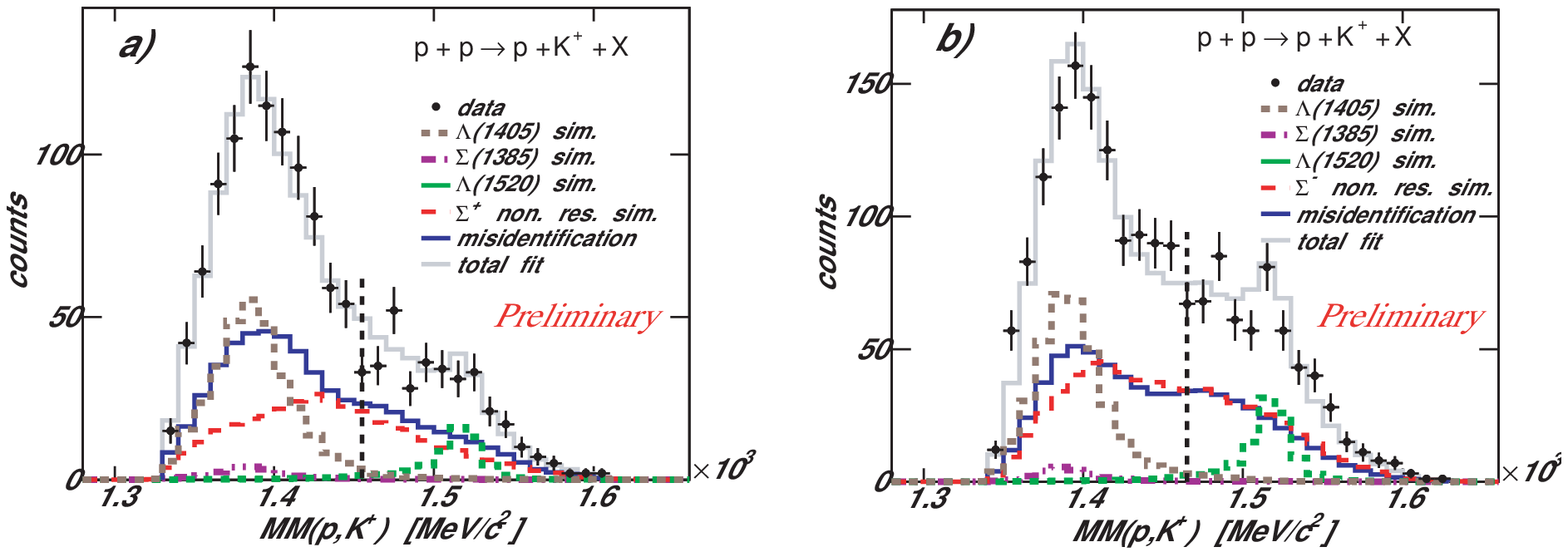}
	\caption{\small{Missing mass of the proton and the K$^{+}$ for the $\Sigma^+$ (a) and $\Sigma^-$ (b) decay channel.}}
	\label{fig:Lambda1405_SigmaPAll_Fit_2}
\end{figure}
Figure \ref{fig:Lambda1405_SigmaPAll_Fit_2} shows the obtained signals (black points) together with the different identified contributions. Besides the $\Lambda(1405)$ and the $\Sigma(1385)^0$, contaminations due to the non resonant $\Sigma$ production and due to $\Lambda(1520)$ were simulated. Additionally, a background attributed to misidentification was modeled with a side band analysis. After subtraction of all background sources and corrections for efficiency and acceptance, the obtained $\Lambda(1405)$ signals show a pole mass at around 1385 MeV/c$^2$, roughly 20 MeV/c$^2$ below the nominal value. This ''mass shift'' might be caused by an attractive interaction between the $\Lambda(1405)$ and the additional produced proton. However, the investigation of the opening angle of these two particles, which should be influenced by such an interaction, showed no significant deviation from phase space simulations. Further studies are underway.

\vfill  

\noindent{\bf References }
\begin{description}
\setlength\itemsep{-3pt}
\item{[1]} T. Yamazaki, Y. Akaishi, Phys. Rev {\bf C65} (2002) 044005. 
\item{[2]} T. Hyodo et al., Mod. Phys. Lett. {\bf A23} (2008) 2393.
\item{[3]} G. Agakishiev et al., Eur. Phys. J. {\bf A41} (2009) 243.
\end{description}


\setcounter{equation}{0} 
\setcounter{figure}{0}
\clearpage

\addcontentsline{toc}{section}{
{\bf The SIDDHARTA experiment -  general overview}\\
D.L.~Sirghi on behalf of the SIDDHARTA Collaboration}


%





\titl{The SIDDHARTA experiment -  general overview}

\name{
D.L.~Sirghi$^{1}$ on behalf of the SIDDHARTA Collaboration
}

\adr{
$^1$  INFN, Laboratori Nazionali di Frascati, CP 13,
 Via E. Fermi 40, I-00044, Frascati (Roma), Italy \\

}


The precision measurement of kaonic atoms at the  DA$\Phi$NE accelerator of the
LNF-INFN Laboratories was performed in the framework of the SIDDHARTA (SIlicon
Drift Detector for Hadronic Atom Research by Timing Application) international
Collaboration. 

The scientific aim of the experiment is to perform a precise measurement of K-series
kaonic hydrogen X-rays and the first ever measurement of the kaonic
deuterium X-rays to determine the strong interaction energy level shifts
and widths of the lowest lying atomic states.

These measurements offer a unique possibility to precisely determine the
isospin dependent antikaon nucleon scattering lengths which are directly
connected with the physics of the kaon nucleon interaction.

\begin{wrapfigure}[20]{r}{8.2cm}
\begin{footnotesize}
\begin{flushleft}
\vspace*{-0.8cm}
\epsfig{file=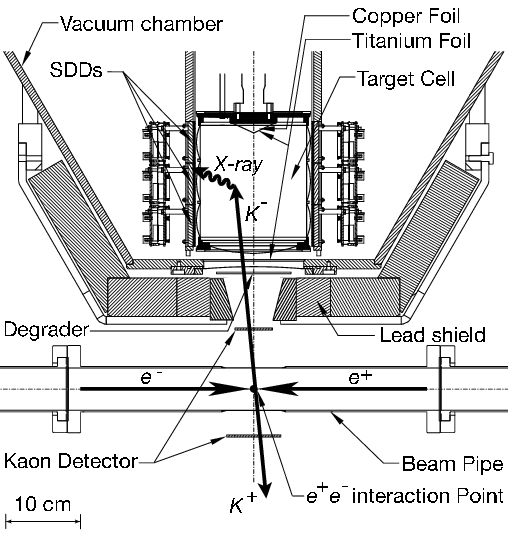,width=8.2cm}
\vspace*{-0.6cm}
\caption{The SIDDHARTA setup.}
\label{setup}
\end{flushleft}
\end{footnotesize}
\end{wrapfigure}

The experiment combined the excellent low energy kaon beam generated at
DA$\Phi$NE, allowing to use gaseous targets, with excellent fast X-rays
detectors, namley the Silicon Drift Detectors. 
An overview of the experimental setup is presented in Fig. \ref{setup}.
The SIDDHARTA experimental setup [1] consists of three main parts: a kaon detector, an X-ray
detection system and a cryogenic target system.


SIDDHARTA was installed on DA$\Phi$NE in
autumn 2008 and took data till late 2009.

In addition to the measurements of kaonic hydrogen and the first ever exploratory 
measurement of the kaonic deuterium, we have
performed the measurements of kaonic helium transitions to the 2{\em p} level
(L-lines): for the first time in a gaseous target for kaonic helium4 [2] and for the
first time ever for kaonic  helium3.

DA$\Phi$NE proved to be an ideal ''kaonic atoms'' factory, representing a unique opportunity to study in a complete way
the kaon nucleon/nuclei physics at low energy.
Presently, data analyses are ongoing. In parallel, plans for an upgrade of the 
experimental setup for an enriched scientific case for the near future are being 
considered

\vfill  

\noindent{\bf References }
\begin{description}
\setlength\itemsep{-3pt}

\item{[1]} J. Zmeskal, Progr. Part. Nucl. Phys. {\bf 61} (2008) 512.
\item{[2]} M. Bazzi, {\it et al.}, Phys. Lett. B {\bf 681} (2009) 310-314
\end{description}


\setcounter{equation}{0} 
\setcounter{figure}{0}
\clearpage

\addcontentsline{toc}{section}{
{\bf Strange multibaryons studied in the $^4$He($K^-_{stopped}, YN$) reaction}\\
T.~Suzuki, for KEK-PS E549 collaboration}

%





\titl{Strange multibaryons studied in the $^4$He($K^-_{stopped}, YN$) reaction}

\name{
T.~Suzuki$^{1}$, for KEK-PS E549 collaboration
}

\adr{
$^1$ Department of Physics, The University of Tokyo, Tokyo 113-0033, Japan \\
}

In order to establish multinucleon absorption processes and find out possible multibaryonic signals from 
non-mesonic branches of stopped $K^-$ reaction on $^4$He,  we have investigated correlations of back-to-back 
coincident  $\Lambda N$, $\Lambda d$ and $\Lambda t$ pairs in the KEK-PS E549 experiment [1, 2, 3]. 

As the first step, we have performed a comprehensive study of multinucleon absorption processes. The following branches have been successfully confirmed as well-separable processes, and the branching ratios have been obtained:
\begin{enumerate}
\item{two-Nucleon Absorption (2NA): }{$\Lambda n/\Lambda p$ branches, from $\Lambda n$ and $\Lambda p$ correlations [4]}
\item{three-Nucleon Absorption (3NA): }{$\Lambda d$ branch, from $\Lambda d$ correlations [3, 5]}
\item{four-Nucleon Absorption (4NA): }{$\Lambda t$ branch, from $\Lambda t$ correlations [5]}
\end{enumerate}

Furthermore, mysterious non-mesonic reaction strengths, which indicate possible signals of the formation and non-mesonic decay of strange multibaryon states with large widths, have been separately identified in $\Lambda N$ spectra. For the $\Lambda p$ spectrum case to which normalized spectrum is given, the branching fraction of the multi-baryonic candidate is larger than that of 2NA, while the 
2NA contribution is more dominant for the $\Lambda n$ case. By looking for one additional neutron, $\Lambda Nn$ coincidence events were extracted from $\Lambda N$ event sets to investigate further the mysrious component, so that $\Sigma^0 pnn$ and $\Lambda pnn$ final states were successfully separated by means of $^4$He$(K^-_{stopped}, \Lambda Nn)$ missing mass. In the reconstructed $\Sigma^0pnn$ final states from $\Lambda p$ events, proton momentum distribution was clearly shifted toward the low-momentum side from the value expected for the 2NA process, and no indication of $\Sigma^0 p$ branch of the 2NA process was obtained. By the fact, we finally confirmed that the source of $\Sigma^0 p$ back-to-back pairs which occupies non-negligible fraction of $\Lambda p$ spectrum, is not due to the two-nucleon absorption process, and the fact almost immediately led the interpretation of the strength as the formation of a strange tri-baryonic state and its $\Sigma^0 nn$ decay.

\vfill  

\noindent{\bf References }
\begin{description}
\setlength\itemsep{-3pt}
\item{[1]} M. Sato {\it et. al}, Phys. Lett. B {\bf 659} (2008) 107.
\item{[2]} H. Yim  {\it et. al}, Phys. Lett. B {\bf 688} (2010) 43.
\item{[3]} T. Suzuki {\it et. al}, Phys. Rev. C {\bf 76} (2007) 068202
\item{[4]} T. Suzuki {\it et. al}, Mod. Phys. Lett. A {\bf 23} (2008) 212303
\item{[5]} T. Suzuki {\it et. al}, ArXiv:1009.5802, to be submitted to Phys. Lett. B
\end{description}


\setcounter{equation}{0} 
\setcounter{figure}{0}
\clearpage

\addcontentsline{toc}{section}{
{\bf $\Lambda^*$-hypernuclei with chiral dynamics}\\
T.~Uchino, T.~Hyodo, M.~Oka}

%





\titl{$\Lambda^*$-hypernuclei with chiral dynamics}

\name{
Toshitaka~Uchino$^{1}$, Tetsuo~Hyodo$^{1}$, Makoto~Oka$^{1}$
}

\adr{
$^1$ Department of Physics, Tokyo Institute of Technology, Meguro 152-8551, Japan
}


The $\bar{K}$-nuclei system is expected to bound deeply, and is actively discussed from theoretical and experimental aspects [1].
Based on several theoretical assertion that the main component of $\bar{K}N$ bound states is $\Lambda^*N$ bound states, for this problem, the $\Lambda^*$-hypernuclei picture where the $\Lambda^*$ is considered as an elementary particle in nuclei is available.
This viewpoint makes it easy to study the few-body $\bar{K}$-nuclei in a similar way to the calculation of normal hypernuclei.
However, the interaction between the $\Lambda^*$ and a nucleon is not clear for the lack of the direct experimental data.
In the previous work, the $\Lambda^*N$ and $\Lambda^*NN$ system are solved with the phenomenological $\Lambda^*N$ one-boson-exchange potential and the $\Lambda^*N$ interaction is determined to reproduce the result of the FINUDA experiment [2].
In the present study, we discuss the lowest energy state of the $\Lambda^*N$ two body system which is the simplest $\Lambda^*$-hypernuclei system by constructing the $\Lambda^*N$ interaction based on the chiral dynamics [3].

The $\Lambda^*N$ system is labeled by two quantum numbers, the orbital angular momentum $L$ and the total spin $S$.
We only consider $s$-wave contribution which is dominant in the lowest energy state, and study for the $S=0$ and $S=1$ case.
To study the bound state of the $\Lambda^* N$ system, we construct the $\Lambda^* N$ one-boson-exchange potential by generalizing the J\"{u}lich (model A) potential.
In order to determine several coupling constants, we adopt the microscopic structure of the $\Lambda^*$ obtained by the chiral unitary approach.
With the chiral unitary approach which is the model based on the chiral dynamics, the $\Lambda^*$ resonance is dynamically generated as a superposition of two states.
Accordingly, the $\Lambda^*N$ system consists of two component, therefore we solve the two channel coupled Schr\"{o}dinger equation.

With parameters in the model [4], we find the bound state solution for the $S=0$ at $M_{\Lambda^*N}=$~2285MeV, while for the $S=1$ case, the physical bound state is not found.
Using the obtained wave function of the bound state, we have the root-mean-square radius of the $\Lambda^*N$ system as $\langle r^2 \rangle=1.10$~fm, which is much smaller than the deuteron.
From the result, the bound state of the $\Lambda^*N$ can be considered as a compact system.
In addition, the decay width can be estimated with wave function perturbatively. The $\pi\Sigma N$ decay width through the fall apart process is obtained as $\Gamma\sim 7$~MeV, where the $\Lambda^*$ inside the bound state decays with the other nucleon being a spectator. Because of the small phase space, the width is relatively narrow. The decay widths to the $\pi\Sigma N$ and $\pi\Lambda N$ channels through the meson-exchange process, as well as the two-body decays into $\Lambda N$ and $\Sigma N$ channels, will be studied elsewhere.

We thank Emiko Hiyama for useful discussion.
This work is partially supported by a grant for the Tokyo Institute of Technology Global COE program, "Nanoscience and Quantum Physics", from the Ministry of Education, Culture, Sports, Science and Technology of Japan.

\vfill  

\noindent{\bf References }
\begin{description}
\setlength\itemsep{-3pt}
\item{[1]} Y.~Akaishi and T.~Yamazaki, Phys.\ Rev.\  C {\bf 65}, 044005 (2002).
\item{[2]} A. Arai, M. Oka and S. Yasui, Prog.\ Theor.\ Phys.\  {\bf 119}, 103 (2008).
\item{[3]} T. Uchino, T. Hyodo and M. Oka, \textit{in preparation}.
\item{[4]} T. Hyodo, S. I. Nam, D. Jido and A. Hosaka, Phys.\ Rev.\  C {\bf 68}, 018201 (2003).
\end{description}


\setcounter{equation}{0} 
\setcounter{figure}{0}
\clearpage

\addcontentsline{toc}{section}{
{\bf Antikaon-Nuclear Few-Body Systems - {\it Theory status} -}\\
W. Weise}

%





\titl{Antikaon-Nuclear Few-Body Systems\\
- {\it Theory status} -}

\name{
W. Weise
}

\begin{center}
\adr{
Physik-Department, Technische Universit\"at M\"unchen \\ D-85747 Garching, Germany
}
\end{center}

An updated overview on low-energy interactions of antikaons with nucleons and nuclei is given [1] summarizing the status of the following topics:\\

1) {\bf Low-energy QCD with strange quarks} is well represented in terms of chiral $SU(3)$ effective field theory with coupled channels [2-4]. Starting from basic principles of spontaneous and explicit breaking of global chiral symmetries in QCD, this approach provides a controlled and well established framework with an effective Lagrangian systematically organized in powers of energy, momentum and quark masses. \\

2) {\bf $\bar{K}N$ threshold physics} provides important quantitative constraints for the chiral $SU(3)$ meson-baryon effective Lagrangian and extrapolations into subthreshold regions. Major improvements are imminent with the new SIDDHARTA kaonic hydrogen data that are expected to set new standards in this context. Current measurements of $\pi\Sigma$ invariant mass spectra using different probes (e.g. photoproduction of $K^+$ at JLab and proton induced reactions with HADES at GSI) are important in order to clarify the nature of the $\Lambda(1405)$ and the underlying coupled-channels dynamics with its two-poles scenario [4].\\

3) Concerning the quest for {\bf $\bar{K}NN$ quasibound systems}, several variational and Faddeev calculations have been performed and reported in the literature, with binding energies ranging between 20 and 80 MeV and widths between 40 and 100 MeV. Substantial differences in these results are primarily reflecting uncertainties in subthreshold (off-shell) extrapolations of the relevant $\bar{K}N$ amplitudes, and to some extent approximations
made in either variational or 3-body Faddeev calculations. An important recent observation [5] concerns the role of the explicit energy dependence of the basic $\bar{K}N$ and $\pi\Sigma$ interactions. This energy dependence directly relates to the chiral symmetry driven leading order couplings of antikaons and pions and their Nambu-Goldstone boson nature. Once the energy dependence is properly incorporated in three-body computations, the two-poles scenario is seen to persist also in the  $\bar{K}NN$ 3-body system [5]. Variational and Faddeev calculations operating with such energy dependent interactions tend to give weak $\bar{K}NN$ binding (with binding energies around 20 MeV) [5,6]. 
\begin{description}
\setlength\itemsep{-3pt}
\item{[1]} W. Weise, Nucl. Phys. {\bf A 835} (2010) 51.
\item{[2]} N. Kaiser, P.B. Siegel, W. Weise, Nucl. Phys. (1995); \\E. Oset, A. Ramos, Nucl. Phys. {\bf A 635} (1998) 99.
\item{[3]} B. Borasoy, R. Nissler, W. Weise, Eur. Phys. J. {\bf A 25} (2005) 79; B. Borasoy, U.-G. Meissner, R. Nissler, Phys. Rev. {\bf C 74} (2006) 055201; R. Nissler, PhD thesis (2008).
\item{[4]} T. Hyodo, W. Weise, Phys. Rev. {\bf C 77} (2008) 035204; \\D. Jido et al., Nucl. Phys. {\bf A 725} (2003) 181. 
\item{[5]} Y. Ikeda, H. Kamano, T. Sato, Prog. Theor. Phys. {\bf 124} (2010) 533.
\item{[6]} A. Dote, T. Hyodo, W. Weise, Nucl. Phys. {\bf A 804} (2008) 197, \\Phys. Rev. {\bf C 79} (2009) 014003.

\end{description}


\setcounter{equation}{0} 
\setcounter{figure}{0}
\clearpage

\addcontentsline{toc}{section}{
{\bf Nuclear states of $\overline{K}$  via ($\overline{K},\gamma$) and $(\overline{K},\pi $) reactions}\\
S. Wycech}

%





\titl{Nuclear states of $\overline{K}$  via ($\overline{K},\gamma$) and $(\overline{K},\pi $) reactions}

\name{
 S. Wycech
}

\begin{center}
\adr{
  Soltan Institute for Nuclear Studies, Warsaw, Poland
}
\end{center}

Chances for detection of the nuclear states of  $\overline{K}$ mesons were  discussed
with  two unconventional methods: ($\overline{K},\pi$) and $(\overline{K},\gamma $) which offer
reasonable branchings of  $ 10^{-4} $ to $10^{-2}$ at the price of high background.

$\bullet$  The  $\overline{K} \rightarrow \gamma $  i.e.  radiative transitions from atomic to 
nuclear states could be noticeable from the atomic P-wave states. In the  hydrogen atoms  such experiments
were undertaken with success [1]. However, the  transition of  interest to our
community -  the  $ p ~ {\bar K} \rightarrow \gamma  ~ \Lambda(1405)$ - which  could provide 
the shape of $\Lambda(1405)$ has not been resolved. One related problem  is   the shape of the
$\gamma$ line. It differs from the Lorentzian.  The rate for this reaction is very sensitive to 
the gas target pressure. With the conditions of SIDDHARTA it may reach  $ 10^{-4} $ per stopped
meson. In helium and deuterium the ratio is higher  but it leads to  less interesting, 
speculative states formed by the presence of $\Sigma(1385)$ in nuclei [2].

$\bullet$   $(\overline{K},\pi). $   This reaction may be interesting in  situations  of 
low momentum transfer  i.e. for very deeply bound $\overline{K} $ meson.  In particular,  the low
energy reaction $^4He, \overline{K} \rightarrow   ( ^4He \overline{K})_{BS} , \pi $ 
may offer a sizable  $\sim 1\% $ branching ratio if the $ ( ^4He \overline{K})_{BS}$ binding energy
is close to 170 MeV. Such values   seem likely in view of the 100 MeV $\overline{K}$NN bindings 
indicated by FINUDA [3] and DISTO [4], it is also indicated by  calculations [2].
It is interesting to note that an old chamber measurement of $^4He, \overline{K} 
\rightarrow   d, p, \Lambda, \pi $ shows emission of very low momentum  $\pi$ mesons not accounted for 
in the standard way [5]. These constitute $\sim 1\% $ of all events and kinematically fit the expected
binding. The calculation of the relevant cross section has one semi-free parameter - the partial 
width for decay of the $(^4He \overline{K})_{BS}$ bound state via the $\overline{K}NN$ non-mesic
capture. The width for such decay mode  was estimated by two
independent groups to be about 10-20 MeV and this leads to a large
$\sim$ mb cross section, consistent with the finding of
ref. [5].

\vspace{1mm}
\noindent

\vfill  

\noindent{\bf References }
\begin{description}
\setlength\itemsep{-3pt}
\item{[1]} D.A. Whitehouse {\it et al.,  \/}, Phys. Rev. Lett. \textbf{63},1352(1989); 
K.D. Larsen et al , Phys. Rev. \textbf{D47},799(1993).

\item{[2]} S. Wycech and A.M. Green, Phys.Rev.\textbf{C79}, 014001 (2009).

\item{[3]} M. Agnello {\it et al}., Phys.Rev.Lett. {\bf 94}, 212303 (2005).

\item{[4]} T. Yamazaki {\it et al.}, Phys.Rev.Lett. \ {\bf 104}, 33502 (2010).

\item{[5]} R. Roosen  {\it et al.},  Nuovo Cim. 49 A, 217(1979).

\end{description}

\setcounter{equation}{0} 
\setcounter{figure}{0}
\clearpage

\addcontentsline{toc}{section}{
{\bf Kaonic Bound States $K^-p$, $K^-pp$, and $K^-K^-pp$: Present and Future}\\
T. Yamazaki}

%





\titl{Kaonic Bound States $K^-p$, $K^-pp$, and $K^-K^-pp$: Present and Future}

\name{Toshimitsu Yamazaki $^1$}

\adr{
$^1$ Department of Physics, University of Tokyo, Bunkyo-ku, Tokyo 113-0033 and RIKEN Nishina Center, Wako-shi, 351-0198 Saitama-ken, Japan \\
}

\noindent
{\bf 1.  $K^- p$: A new proposal} 

~~~To solve the current debate on the position of the quasi-bound $K^-p$ state, we propose to measure the   $\Sigma\pi$ invariant-mass spectrum in stopped-$K^-$ absorption in deuteron, since the spectrum, reflecting the soft and hard deuteron momentum distribution, is expected to have a narrow quasi-free component with an upper edge of $M = 1430$ MeV/$c^2$, followed by a significant ``high-momentum" tail toward the lower mass region, where a resonant formation of $\Lambda(1405)$ of any mass and width in a wide range will be clearly revealed.  [1]\\

\noindent
{\bf 2.  $K^- pp$: Additional DISTO information}

~~~The DISTO data on the exclusive $pp \rightarrow p \Lambda K^+$ reaction at 2.85 GeV revealed a broad distinct peak of 26-$\sigma$ confidence 
 with a mass $M_X = 2265 \pm 2~(stat) \pm 5~(syst)$ MeV/$c^2$ and a width $\Gamma_X = 118 \pm 8~(stat) \pm 10~(syst)$ MeV. The enormously large cross section indicates formation of a compact $K^-pp$ with a large binding energy of $B_K = 105$ MeV, which can be a possible gateway toward cold and dense kaonic nuclear matter [2].  \\
 
\noindent
{\bf 3. $K^-K^-pp$: A new proposal}

~~~The idea to form $K^-pp$ is now extended to form $K^-K^-pp$ in high-energy $pp$ collision.  
\begin{figure}[h]
\centering
\includegraphics[height=.35\textheight]{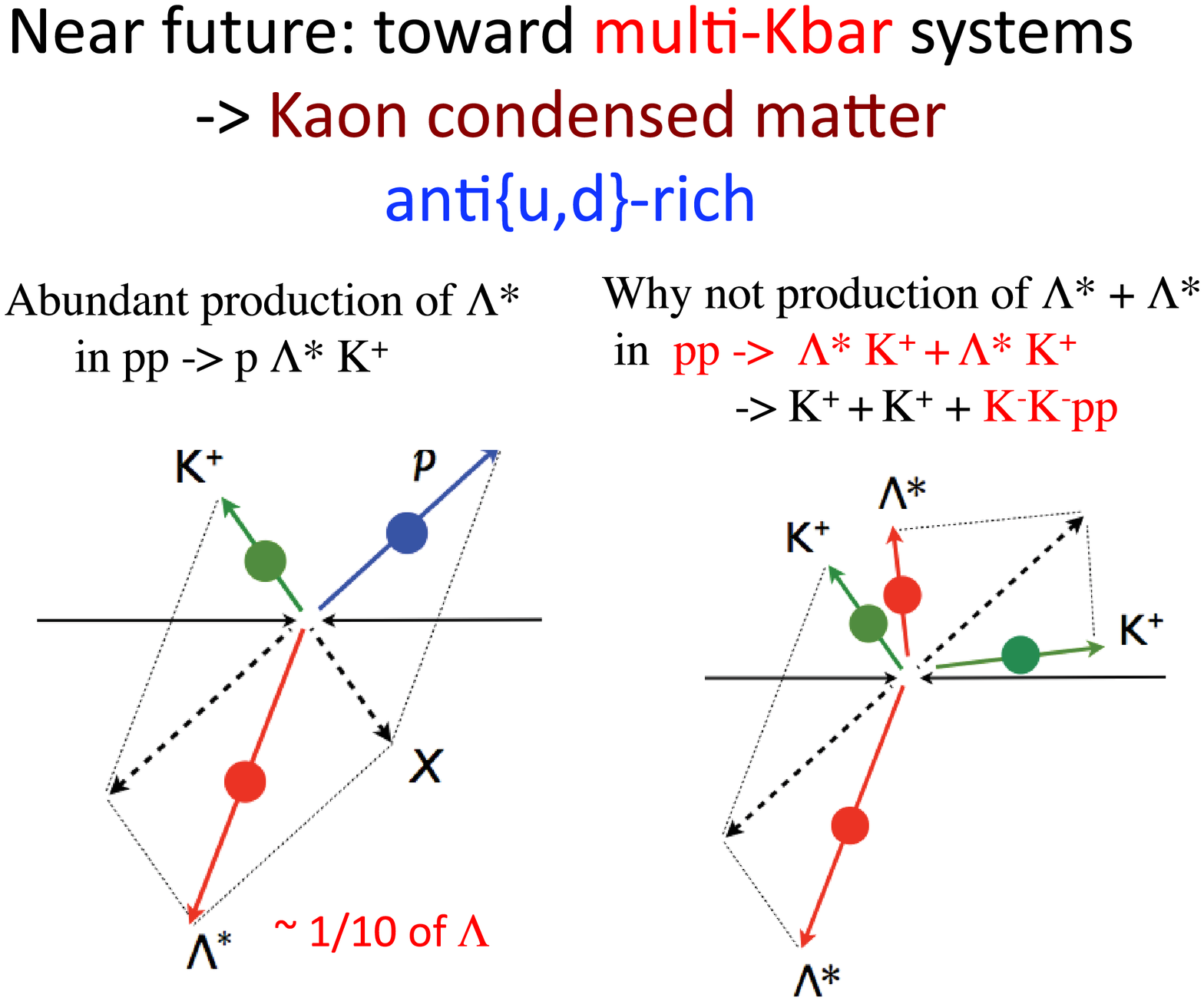}
\end{figure}

\vfill  

\noindent{\bf References }
\begin{description}
\setlength\itemsep{-3pt}
\item{[1]}  J. Esmaili, Y. Akaishi and T. Yamazaki, arXiv:0909.2573v3; Phys. Lett. B686 (2010) 23.
\item{[2]}  T. Yamazaki {\it etal.}, Phys. Rev. Lett. 104 (2010) 132502.
\end{description}


\setcounter{equation}{0} 
\setcounter{figure}{0}
\clearpage


\addcontentsline{toc}{section}{
{\bf Conference Photos} }

 
 
 

\titl{Conference Photos}

\vspace{5cm}

\begin{figure}[htbp]
\begin{center}
\includegraphics[width=12cm]{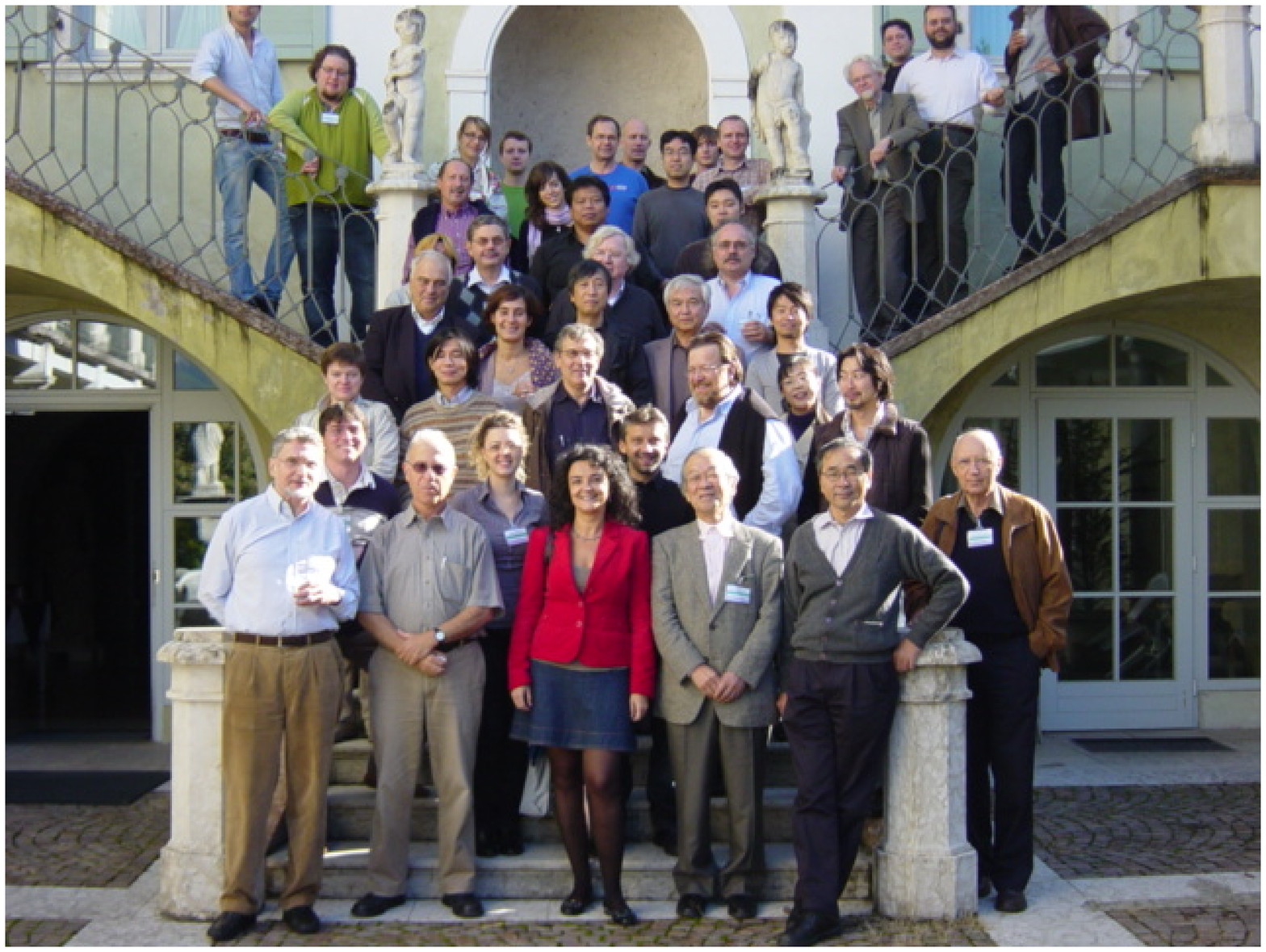}
\end{center}
\end{figure}


\setcounter{equation}{0} 
\setcounter{figure}{0}
\clearpage

\end{document}